\documentclass[12pt,a4paper]{article}
 \pdfoutput=1

\usepackage{jheppub}
\usepackage{physics}
\usepackage{relsize}
\usepackage{tikz}
\usepackage{pgfplots}
\usepackage{comment}
\usepackage{ulem} 
\usepackage[utf8]{inputenc}

\newcommand{\be}{\begin{equation}}
\newcommand{\ee}{\end{equation}}

\title{A substrate for brane shells from $T\bar{T}$}
\author[a]{Jeremias Aguilera-Damia,}
\author[b]{Louise M. Anderson,}
\author[b]{Evan Coleman,}
\affiliation[a]{Centro At\'omico Bariloche, CNEA and CONICET, Bariloche, R8402AGP, Argentina}
\affiliation[b]{Stanford Institute for Theoretical Physics and Department of Physics, Stanford University, Stanford, CA 94305, USA}

\emailAdd{jeremiasadlp@gmail.com}
\emailAdd{louise.m.a.anderson@gmail.com}
\emailAdd{ecol@stanford.edu}

\begin{document}

\abstract{
 A solvable current-current deformation of the worldsheet theory of strings on $AdS_3$ has been recently conjectured to be dual to an irrelevant deformation of the spacetime orbifold CFT, commonly referred to as single-trace $T\bar T$. These deformations give rise to a family of bulk geometries which realize a non-trivial flow towards the UV. For a particular sign of this deformation, the corresponding three-dimensional geometry approaches $AdS_3$ in the interior, but has a curvature singularity at finite radius, beyond which there are closed timelike curves. It has been suggested that this singularity is due to the presence of ``negative branes,'' which are exotic objects that generically change the metric signature. We propose an alternative UV-completion for these geometries by cutting and gluing to a regular background which approaches a linear dilaton vacuum in the UV. In the S-dual picture, a singularity resolution mechanism known as the enhan\c{c}on induces this transition by the formation of a shell of $D5$-branes at a fixed radial position near the singularity. The solutions involving negative branes gain a new interpretation in this context.

}

\maketitle

\section{Introduction}
\label{sec:Introduction}

The $T\bar{T}$ deformation of two-dimensional quantum field theories (QFTs) has recently attracted attention in a diverse range of physics subfields, due to its universality and solvability. Its universal nature stems from the fact that the double-trace operator defining the deformation is made out of products of the stress tensor,
\begin{align} \label{eqn:TTbarDef}
        T\bar{T}\equiv T^{\mu\nu}T_{\mu\nu} - (T^\mu_\mu)^2\, .
    \end{align}
It is a nontrivial result that this operator is free of short-distance singularities, independently of the details of the local QFT \cite{Zamolodchikov:2004ce}. 
Even though the deforming operator is irrelevant in the renormalization group sense, it triggers a flow which is solvable toward the UV. This flow is usually parametrized by a coupling $\mu$ of dimensions of length squared:
 \begin{align} \label{eqn:TTbarFlow}
        \dfrac{\partial \log{Z}}{\partial \mu} = \int d^2x\, \langle T\bar{T} \rangle_\mu.
    \end{align}
Note that $\langle \cdot \rangle_\mu$ denotes the expectation value computed in the theory at each point along the flow. Its solvability depends on the fact that some important observables can be computed exactly as functions of the $T\bar T$ coupling $\mu$ \cite{Smirnov_2017,Cavaglia:2016oda,Cardy:2018sdv,Cardy:2019qao,Donnelly:2018bef}.  An important feature of these flows is that the resulting dynamics in the UV are strongly dependent on the sign of $\mu$. For positive coupling, the theory becomes non-local, displaying a minimal length and Hagedorn growth for the high energy density of states. On the contrary, for negative coupling, the spectrum develops complex energy levels at a given scale, and a suitable UV completion is still an open question. See \cite{Jiang:2019hxb} for a pedagogical review.

This deformation has opened new avenues in the study of quantum gravity. For example, one can derive the $T\bar T$ flow by coupling the undeformed theory to topological gravity in two dimensions \cite{Dubovsky:2012wk,Dubovsky:2017cnj,Dubovsky:2018bmo,Coleman:2019dvf,Tolley:2019nmm,Mazenc:2019cfg}. In the context of AdS/CFT, it has an interpretation as implementing Dirichlet boundary conditions at finite radius in the bulk~\cite{MMV,Kraus:2018xrn}, leading to explicit implementations of de Sitter holography \cite{Gorbenko:2018oov,Lewkowycz:2019xse}. These interesting bottom-up constructions so far apply exclusively to the gravitational sector, whereas the inclusion of general bulk matter is not yet fully understood (see {\it e.g.}~\cite{Hartman:2018tkw} for extensions in this direction). Furthermore, these holographic realizations are well-suited for negative values of the coupling, where the complex energy levels naturally arise as a consequence of the presence of a Dirichlet wall \cite{MMV}. Recently, similar proposals implementing mixed boundary conditions at spatial infinity and conformal boundary conditions have been shown to overcome some of these abnormalities~\cite{guica2019tbar,Coleman:2020jte}.

It is therefore worth exploring these flows through the lens of more complete holographic realizations. String theory on $AdS_3$ backgrounds supported by Neveu-Schwartz (NS) flux is currently one of the best understood constructions in quantum gravity \cite{Giveon:1998ns,Kutasov:1999xu}. There has been great progress toward a concrete top-down realization of holography in these backgrounds~ \cite{Seiberg:1999xz,Argurio:2000tb,Eberhardt:2018ouy,Eberhardt:2019ywk,Gaberdiel:2020ycd}. On a similar note, a non-gravitational decoupling limit of String Theory in the presence of NS fluxes is suitably accounted for by Little String Theory (LST), which was shown to accurately describe the low energy dynamics which take place in the world volume of $NS5$-branes \cite{Aharony:1998ub,Witten:1997kz}. Such exotic theories are generically non-local and display Hagedorn growth at high energies~\cite{Giveon:2005mi}. 

These features inspired a novel realization of the $T\bar T$ flow, referred to as ``single-trace'' $T\bar T$. This new flow is implemented holographically by an exactly marginal current-current deformation of the worldsheet sigma model of strings in $AdS_3$ with NS fluxes \cite{Giveon:2017nie,Giveon:2017myj,Asrat:2017tzd}. Here, the role of the dimensionful coupling $\mu$ is played by the squared string length $\alpha'$. As its name suggests (and in contrast to its double-trace counterpart), the trajectory triggered by the single-trace $T\bar T$ deformation involves strong backreaction on the background where the strings live. We will briefly review some aspects of these constructions in section \ref{sec:JJbarIntro}. As before, the fate of the theory in the UV strongly depends on the sign of the deformation parameter. For positive coupling, the resulting backgrounds  interpolate smoothly between $AdS_3$ and a linear dilaton vacuum of LST, naturally implementing the Hagedorn growth. On the contrary, the flow towards negative couplings may lead to severe violations of causality in the target spacetime, manifest in curvature and dilaton singularities, usually accompanied by the development of closed timelike curves (CTCs). The singular case will be the main focus of this article. 
 
In~\cite{Chakraborty:2020swe}, the authors noted that such signature changes are generic to exotic ``negative branes''~\cite{dijkgraaf2016negative}, and they demonstrated that the impact of the singularity on certain stringy probes was benign. Fundamental strings pass through it without issue.\footnote{However, they still encounter an analogous energy cutoff as they travel beyond the singularity \cite{Chakraborty:2020swe,Chakraborty:2020cgo}.} Moreover, if the vacuum is excited so as to contain an event horizon, then as the horizon is taken to reach the singularity, they will begin to expand together due to backreaction. These constructions correctly account for the main phenomenological signatures of the flow towards negative coupling, featuring in particular a maximal energy state.

However, it is important to consider whether a string-theoretic resolution of the singularity exists which avoids the introduction of exotic objects. Such a resolution would ideally excise the region with CTC's, as has been shown to occur in similar settings via the condensation of winding tachyons and light winding strings~\cite{McGreevy_2005,Costa_2005}. A tantalizing possibility would amount to a string compactification, providing a UV-complete analogue of the Dirichlet wall. The deformation derived in~\cite{Gorbenko:2018oov}\ (and generalized in~\cite{Lewkowycz:2019xse}) is also available in the single-trace case, and that might be the most natural setting for this question, since de Sitter is compact.  In this work, we will not address that question, but will find a new trajectory related to the single-trace one that agrees with the deformed geometry in the IR, but crosses over to the other sign in the UV. This trajectory is non-singular and free of exotic objects.

We must clarify what ``resolve the singularity'' means in this context. It is of broad interest to understand whether string theory is a finite and UV-complete theory of quantum gravity, with supergravity  as its low-energy effective description. However, supergravity backgrounds generically exhibit singularities, e.g. in the Ricci curvature and the string coupling. These divergences present a challenge to the finiteness of the UV-complete theory, and the onus is on us to demonstrate how stringy corrections prevent the formation of these sicknesses. To this end, a program arose to classify the species of singularities which can appear in supergravity backgrounds, and show how they are excised by stringy physics. So far, a few major classes of singularities have been identified, each with their own resolution mechanisms: orbifolds~\cite{DIXON1985678}, conifolds~\cite{Witten:1993yc,Aspinwall:1993yb}, flops~\cite{Strominger:1995cz,Greene:1995hu}, and repulsons~\cite{Repulson1,Repulson2,Repulson3}. 

Herein, we propose a possible resolution which is related by S-duality to the mechanism applied to resolve certain repulson singularities, the ``enhan\c{c}on.'' Our guiding principle will be to obtain a regular background, while maintaining the resemblance to single trace $T\bar T$ backgrounds in the IR (and, incidentally, in the UV). Given that, it is important to remark that our proposal is not an alternative nor a correction to the singular flow described in \cite{Chakraborty:2020swe}, but it should correspond to a completely different trajectory, involving other deformations as the energy scale increases. In principle, there may be other valid answers to this question, as taking a singular supergravity background to its UV completion will generically give a one-to-many correspondence. Nevertheless, some of the degeneracy can be reduced by physical considerations. Here, we develop what we believe to be the simplest picture amongst the generalizations which are readily available. 
\\
\\
\textbf{Organization of the Paper}
\\
This article is organized as follows. First, in section~\ref{sec:JJbarIntro}, we review the single-trace version of the $T\bar{T}$ flow. In section~\ref{sec:3DPicture}, we restrict to a three-dimensional effective theory and rederive the singular geometries of interest, explaining the equivalence to ~\cite{Chakraborty:2019mdf}. We identify their problematic features, and derive a regular solution by cutting and gluing to a regular geometry. This simplified setup is intended as preparation for section~\ref{sec:10DPicture}, where we embed the resolution into a ten-dimensional picture. Working with the S-dual configuration, we show that, for a particular choice of 4-manifold in the compactification, the physics behind the resolution is closely related to the enhan\c{c}on mechanism. This finding leads us to consider a resolution of the singularity by a shell of fivebranes wrapped on a $K3$ manifold. In section~\ref{sec:comments}, we discuss some implications and possible interpretations for the dual QFT. We finish with some concluding remarks and future directions in section~\ref{sec:Discussion}.

\section{A holographic realization of single-trace $T\bar T$}
\label{sec:JJbarIntro}

We begin by briefly reviewing the single-trace $T\bar T$ deformation developed in~\cite{Giveon:2017nie}. The model starts from studying string theory with $AdS_3$ backgrounds supported by NS fluxes, which are determined by two integer charges $Q_1$ and $Q_5$. The worldsheet theory is known to be described by an $SL(2,{\mathbb R})$ Wess-Zumino-Witten (WZW) model with left and right moving current algebras at level $Q_5$ \cite{Giveon:1998ns,Kutasov:1999xu,Maldacena:2000hw}. The $AdS_3$ radius is $R_{AdS}^2=\sqrt{Q_5}\alpha'$. We will work in the regime of $Q_1>Q_5>1$ to ensure both small curvature and weak coupling.

States in the worldsheet theory are classified in terms of representations of the $SL(2,{\mathbb R})_{Q_5}$ algebra \cite{Maldacena:2000hw}. In particular, excitations above the R vacuum belong to the continuous series representations and have vanishing gap in the large $Q_5$ regime. This is the so-called long string sector. This sector is 
conjectured to be dual to a spacetime symmetric product CFT~\cite{Argurio:2000tb,Eberhardt:2018ouy,Eberhardt:2019ywk,Gaberdiel:2020ycd} of the form ${\cal M}^{Q_1}/S_{Q_1}$, with ${\cal M}$ a compact CFT of central charge $c_{\cal M}= 6Q_5$. The total central charge then goes as $c=6Q_1Q_5$. We will focus on the regime of large $Q_1$, for which the holographic realization described so far is amenable to an analysis by perturbative methods in string theory.

A particularly interesting deformation is obtained in this context by the inclusion of a marginal current-current operator in the worldsheet theory
\be
\frac{\partial S_{ws}}{\partial \lambda}= \int d^2z\, J^- \bar J^-,
\label{eq:JJ}
\ee
with $J^-$ ($\bar J^-$) the holomorphic (antiholomorphic) current corresponding to the spacetime Virasoro $L^{-1}$ ($\bar L^{-1}$) and $\lambda$ the dimensionless coupling measuring the strength of the deformation.\footnote{ A family of exactly marginal deformations of the form \eqref{eq:JJ} have been recasted as an $O(d,d)$ transformation of the sigma model in \cite{Hassan:1992gi}. This point of view has been recently explored in the context of solvable irrelevant deformations in \cite{Araujo:2018rho}.} The resulting sigma model is solvable in the sense that, by integrating out certain auxiliary fields, a string theory background can be obtained for any value of the deformation parameter. The resulting metric, dilaton, and 2-form flux read\footnote{Models of this sort which interpolate between two decoupling regimes of $F1$-$NS5$ (or alternatively $D1$-$D5$) systems have been already identified with marginal current-current deformations some time ago, {\it c.f.} \cite{Israel:2003ry} and references therein. It is however in~\cite{Giveon:2017nie} where the deforming operator is connected to the single-trace variant of $T\bar T$.}
\begin{align}
\frac{ds^2}{\alpha'} &= Q_5  dy^2+a\,d\gamma d\bar\gamma \label{eq:bgnd1} \\[0.5em]
a&=\frac{e^{2y}}{1+\lambda e^{2y}} \label{eq:bgnd2}\\[0.5em]
\frac{e^{2\Phi}}{v}&= \frac{ Q_5}{Q_1}\frac{1}{1+\lambda e^{2y}} \label{eq:bgnd3}\\[0.5em]
B&=\frac12\frac{e^{2y}}{1+\lambda e^{2y}}d\bar\gamma \wedge d\gamma. \label{eq:bgnd4}
\end{align}
In the above parametrization, $y\in (-\infty,\infty)$ plays the role of a spacetime radial direction, with holomorphic (antiholomorphic) spacetime coordinates denoted by $\gamma$ ($\bar \gamma$). The dimensionless parameter $v$ denotes a 4-volume scale relating to the string theory compactification from which this background can be obtained.

For positive $\lambda$, the above background interpolates between the Poincar\'e patch of $AdS_3$ at $y\to -\infty$ and a linear dilaton vacuum of LST at $y\to+\infty$. The transition point between these regimes is given by $e^{2y}\sim 1/\lambda$. Note that both regimes are weakly coupled as long as $Q_1$ is sufficiently large.

These two regimes have well-defined holographic duals. As explained above, the infrared $AdS_3$ is dual to a symmetric product CFT in 2-dimensions. On the other hand, the ultraviolet LST vacuum corresponds to the holographic dual of a given non-local, non-gravitational theory (associated to the worldvolume of $NS5$-branes) which is known to present Hagedorn growth at high energies~\cite{Aharony:1998ub,Giveon:2005mi}.

The background \eqref{eq:bgnd1}-\eqref{eq:bgnd4} therefore stands as a holographic realization of a nontrivial renormalization group flow between an IR CFT and a non-local theory with Hagedorn growth in the UV. This remarkable feature has since been connected to the flow driven by the irrelevant $T\bar T$ deformation~\cite{Giveon:2017nie}. Crucially, the flow which describes the interpolation is not the usual notion of $T\bar T$ flow as in~(\ref{eqn:TTbarFlow}), which is controlled by the double-trace operator in (\ref{eqn:TTbarDef}). It is instead a single-trace variant, which applies a $T\bar{T}$ deformation to each $\mathcal{M}$ in the symmetric product, and shares many properties with the traditional $T\bar T$ flow. More precisely, the spacetime theory along the flow is of the form
\be
{\cal M}_{T\bar T}^{Q_1}\Big/S_{Q_1}
\label{eq:defCFT}
\ee 
where ${\cal M}_{T\bar T}$ denotes the deformation by the irrelevant double-trace $T\bar T$ acting on a single factor of the symmetric product.

Backgrounds of this sort can be obtained from compactification of type IIB string theory vacua of the form $AdS_3\times {\cal N}$, with ${\cal N}$ some compact CFT whose central charge is determined by criticality.
Of particular interest are the cases which arise by adding $Q_1$ fundamental strings to a linear dilaton geometry of the form ${\mathbb R}^{1,1}\times S^1\times S^3\times {\cal M}_4$ corresponding to the near-horizon region of $Q_5$ $NS5$-branes wrapped on $S^1\times {\cal M}_4$. The strings are stretched on $S^1$, leading to a BPS configuration whose near-horizon limit corresponds to $AdS_3\times S^3\times {\cal M}_4$.  Here ${\cal M}_4$  denotes a complex dimension 2 Calabi-Yau manifold which can be taken to be either $T^4$ or $K3$. 

In this context, there is a natural interpretation of the parameter $\lambda$ in terms of the squared string length $\alpha'=\ell_s^2$ and the size $R$ of the $S^1$: $\lambda = \alpha'/R^2$. To derive this relationship, one studies the perturbative spectrum of long string states, with energies $E\ll Q_1/R$.  
From the holographic perspective, the effective theory associated to a single string corresponds to a single factor in \eqref{eq:defCFT} and, for states within the aforementioned perturbative regime, each factor is effectively decoupled from the rest. Imposing the Virasoro constraints for the untwisted sector of long strings associated with these states leads to~\cite{Giveon:2017nie,Chakraborty:2019mdf} 
\begin{equation}
\left(E+\frac{R}{\alpha'}\right)^2+\left(\frac{R}{\alpha'}\right)^2 = \frac{2R}{\alpha'}E_0 + \left(\frac{n}{R}\right)^2
\label{eq:defenergy}
\end{equation}
where $E_0= h+\bar h-\frac{Q_5}{2}$ denotes the eigenvalue of the spacetime $L_0+\bar L_0$ in the undeformed IR $AdS_3$ and $n$ measures the momentum along the $S^1$. Solving \eqref{eq:defenergy} leads precisely to the $T\bar T$ spectrum of~\cite{Smirnov_2017,Cavaglia:2016oda} upon the following identification of the dimensionless parameter\footnote{Note the standard dimensionful coupling $\mu$ associated to the $T\bar T$ deformation of the dual CFT is then $\sim \alpha'$.}
\be
\lambda = \frac{\alpha'}{R^2}
\label{eq:lambda}
\ee
The same relation is found when we study the spectrum of high energy states, $E\gg Q_1/R$, which lies in the non-perturbative regime of the theory and is thus described by black holes \cite{Giveon:2005mi,Chakraborty:2020swe}.

The extremal background associated to this $F1$-$NS5$ configuration, which preserves eight of the supersymmetries of type IIB supergravity, can be written in the following form:
\begin{align}
ds^2 &= f_1^{-1}\left(-dt^2+dx^2\right)+ f_5 \left(dr^2+r^2 d\Omega_3^2\right)+V^{1/2}ds_{{\cal M}_4}^2 \label{eq:10d+metric}\\
e^{2\Phi}&= g_s \frac{f_5}{f_1}\label{eq:10d+dilaton}\\
H_3&= d\left(\frac{g_s^2\alpha' Q_1}{v r^2 f_1}\right)\wedge dx\wedge dt + 2 \alpha' Q_5  \epsilon_{3} \label{eq:10d+H3}
\end{align}
where $d\Omega_3^2$ is the metric of the unit 3-sphere, $\epsilon_3$ its associated volume form, and $V = (2\pi)^4\alpha'^2 v$ denotes the asymptotic volume of the compact manifold ${\cal M}_4$ (note that we have not specified the particular manifold yet). The integers $Q_1$ and $Q_5$ measure the NS flux (in units of string length $\ell_s = \sqrt{\alpha'}$), and the harmonic functions $f_1$, $f_5$ read
\begin{align}
f_1= 1+ \frac{r_1^2}{r^2} \,\,\, &, \,\,\, f_5 = 1+\frac{r_5^2}{r^2} \\
r_1^2 = \frac{g_s^2 \alpha' Q_1}{v} \,\,\, &, \,\,\, r_5^2 = \alpha' Q_5
\end{align}
 
The connection between the above background and the 3-dimensional one presented in \eqref{eq:bgnd1}-\eqref{eq:bgnd2} is achieved by performing the LST decoupling limit. This amounts to taking $g_s\to 0$, thus effectively decoupling the gravitational modes from the branes, while focusing on length scales of order $g_s\sqrt{\alpha'}$. Since this limit plays an important role in our results, let us be precise about its implementation. We introduce the coordinate 
    \begin{equation}
        u^2 = \frac{r^2}{g_s^2\alpha'}. 
    \end{equation}
After taking $g_s\to 0$, we find 
    \begin{align}
        ds^2 &= f_1^{-1}\left(-dt^2+dx^2\right)+ Q_5 \alpha' \left(\frac{du^2}{u^2}+ d\Omega_3^2\right)+V^{1/2}ds_{{\cal M}_4}^2 \\
        e^{2\Phi}&= \frac{Q_5}{u^2 f_1}.
    \end{align}
Note in particular that the 3-sphere decouples from the rest of the geometry. Crucially, the $f_1$ harmonic function retains its form, now written in terms of the coordinate $u$. For small enough $u$, we recover the $AdS_3$ throat in the IR regime. 
Finally, after compactifying on $S^3\times {\cal M}_4$ and defining
    \be
        \frac{v u^2}{Q_1} =  \lambda e^{2y} \,\,\, , \,\,\, \gamma= \frac{x-t}{R} \,\,\, , \,\,\, \bar\gamma= \frac{x+t}{R}
    \ee
with $\lambda =  \frac{\alpha'}{R^2}$, the resulting 3-dimensional background reproduces \eqref{eq:bgnd1}-\eqref{eq:bgnd4}. 

So far this construction makes sense for positive values of $\lambda$. Given the definition of $\lambda$ above, having negative $\lambda$ would amount to considering imaginary length scales. Nevertheless, we can analytically continue the background to account for negative values of the coupling. The extremal solution which results is of the same form as \eqref{eq:10d+metric}-\eqref{eq:10d+H3} but with a different harmonic function $f_1$, which now reads
    \begin{equation}
        f_1 = -1+\frac{r_1^2}{r^2}.
        \label{eq:10-f1}
    \end{equation}
The physical meaning of such a solution is obscured by the presence of a naked singularity occurring at $r=r_1$, with CTCs in the exterior region $r>r_1$. A possible interpretation has been proposed in~\cite{Chakraborty:2020swe}, where the singular behavior and signature change is associated to the presence of negative (often called ``ghost'') strings. Such objects come from a family of unconventional negative tension objects in string theory~\cite{dijkgraaf2016negative}. A consistent treatment of black hole configurations corresponding to negative black strings has been considered in~\cite{Chakraborty:2020swe}. The high-energy spectrum obtained is consistent with the spectrum of a $T\bar T$-deformed theory, with the sign of the coupling $\mu$ for which the theory has a UV cutoff. 

It is natural to ask whether we can embed these geometries into more conventional, non-singular backgrounds constructed out of standard string theory objects. Seeking such an embedding will be the goal of the rest of this paper. 

\section{Singularity resolution in three dimensions}
\label{sec:3DPicture}

As described in the previous section, backgrounds of the form \eqref{eq:bgnd1}-\eqref{eq:bgnd4} can be obtained from compactifications of 10-dimensional type IIB string theory with Neveu-Schwartz fluxes, whose action in string frame reads
    \begin{equation}
        S= \frac{1}{16\pi G_N^{(10)}}\int dx^{10}\sqrt{-g} e^{-2\Phi}\left(R+4(\partial \Phi)^2-\frac{1}{12}H^2\right)
    \end{equation}
with $G_N^{(10)}$ the 10-dimensional Newton's constant given by
    \begin{equation}
        G_{N}^{(10)} = 8\pi^6 g_s^2 \alpha'^4.
    \end{equation}
Note that we are absorbing the bare string coupling $g_s$ into the gravitational coupling. 

We look for solutions of the form $M_3\times S^3\times {\cal M}_4$, with $M_3$ some non-compact 3-manifold and ${\cal M}_4$ a complex dimension 2 manifold which can be taken to be either $T^4$ or $K3$. The ansatz for the string frame metric takes the following form
    \begin{equation}
        ds^2=e^{2 D}g_{\mu\nu}dx^\mu dx^\nu +e^{2L}d\Omega_3^2 + e^{2\tilde V}ds^2_{{\cal M}_4}
        \label{eq:3dansatz}
    \end{equation}
where $g_{\mu\nu}$ stands for the Einstein frame metric in $M_3$ and we have parametrized the volumes of the compact submanifolds as
    \begin{equation}
        {\rm Vol}\left(S^3\right) = 2\pi^2 \alpha'^{3/2}e^{3L} 
        \,\,\, , \,\,\,
        {\rm Vol}\left({\cal M}_4\right) = (2\pi)^4 \alpha'^2 e^{4\tilde V}.
    \end{equation}
Note that, in terms of the $v$ parameter introduced earlier, we have $v= e^{4\tilde V}$. The 3-dimensional dilaton field $D$ is of the form
\be
e^{2 D} = e^{2(2\Phi-4\tilde{V}-3 L)}
\label{eq:3ddil}
.\ee
We also include electric and magnetic fluxes arising from the fundamental strings and the $NS5$-branes (respectively), which satisfy  
\be
 \frac{1}{4\pi^2\alpha'}\int e^{-2\phi} * H = Q_1
\,\,\, , \,\,\,
\frac{1}{4\pi^2\alpha'}\int H = Q_5 
\ee
for some integers $Q_1$, $Q_5$.

The resulting effective action for the 3-dimensional Einstein frame metric and moduli is\footnote{Note the scalar fields are not canonically normalized.}
    \begin{equation}
        S=\frac{1}{16\pi G_N^{(3)}}\int d^3 x \sqrt{g} \left(R  -(\partial D)^2-3(\partial L)^2-4(\partial \tilde V)^2  -{\cal V}\right),
        \label{eq:3daction}
    \end{equation}
where the bare 3-dimensional Newton's constant reads
\be
G_{N}^{(3)}= \frac{G_N^{(10)}}{32\pi^6 \alpha'^{7/2}}
.\ee
The effective potential ${\cal V}$ accounts for the effect of the fluxes, together with the contribution from the positive curvature of the $S^3$, and takes the form
\be
{\cal V} = -6 e^{2 D-2L}+2Q_5^2e^{2D-6L}+2Q_1^2 e^{4 D} 
\label{eq:3dpotential}
.\ee 
Note that ${\cal V}$ is independent of $\tilde V$, so we will fix that modulus to an arbitrary constant. The particular value of this constant will not be relevant in the following discussion, so we will ignore it for now, but it will be reintroduced as a parameter when studying the 10-dimensional realization in section \ref{sec:10DPicture}.

It can be readily checked that the following is a solution for the equations of motion of the effective action \eqref{eq:3daction} for any value of $\lambda$
\begin{align}
ds^2 &= e^{-2D_0}\left(Q_5 dy^2+a_0 d\gamma d\bar\gamma\right) \nonumber \\
e^{2D_0}&= \frac{1}{Q_5 Q_1^2}\left(\frac{1}{1-\lambda e^{2y}}\right)^2 \label{eq:sol0}\\ 
a_0 &= \frac{e^{2y}}{1-\lambda e^{2y}} \,\,\, , \,\,\, e^{2L_0}= Q_5 \nonumber
.\end{align}
Note that the above solution is no more than the $\lambda\to-\lambda$ continuation of \eqref{eq:bgnd1}-\eqref{eq:bgnd3}. More precisely, it is the solution arising in the decoupling limit of the background \eqref{eq:10d+metric}, \eqref{eq:10d+dilaton} with the harmonic $f_1$ given by \eqref{eq:10-f1}.

However, the above solution develops a naked singularity at $y=y_s$ with
\be
 e^{2y_s} = \frac{1}{\lambda }, \label{eq:3dsing}
\ee
and the solution ceases to be valid there. We then need to look for an alternative geometry which resolves this singular behavior. There are two additional conditions we want these regular solutions to satisfy: (i) they must solve the same equations of motion in the bulk on a finite region including the origin. That is, we want to keep the same values of $Q_1$ and $Q_5$ in the IR, as they determine the central charge of the dual (undeformed) CFT; (ii) along the same line of reasoning, they must approximate \eqref{eq:sol0} for $-\infty<y<y_{IR}$ for some scale $y_{IR}$. These two conditions ensure that the new solutions will obey the same qualitative behavior as the $T\bar T$-deformed backgrounds in an IR region sufficiently far away from the naked singularity. 

After imposing the above conditions, not much freedom remains. Below we will consider a model that fulfills the above requirements and resolves the singular behavior. We construct it by inserting non-trivial boundary conditions at an incision radius $y_i<y_s$, so that for $y<y_i$ the ``interior'' geometry satisfies the constraints described above. Across the boundary at $y=y_i$, the geometry is glued to what we call the ``exterior'' geometry. It is a solution to modified equations of motion, with shifted $Q_1$ and $Q_5$. In 10 dimensions, the resulting non-singular family of solutions can be connected to a known singularity resolution mechanism in string theory, the so-called ``enhan\c{c}on'' mechanism, as we will show in section~\ref{sec:10DPicture}.

\subsection{Boundary conditions at fixed radius}

To implement the boundary conditions, we insert extended 1- and 5-dimensional objects at $y=y_i$. These objects might be seen as {\it e.g.} fundamental strings, $NS5$-branes, or orientifold planes. From the point of view of the 3-dimensional effective theory, such an interpretation is not necessary.

We configure the boundary sources to be parallel to the original brane configurations to which these geometries are associated. The 1- and 5-dimensional hypersurfaces at the interface will be wrapped over $S^1$ and $S^1\times {\cal M}_4$, respectively. Their tensions in Einstein frame can be deduced by requiring that they match the string frame tension: 
\be
\sigma_s \sqrt{g_{str}} \delta(y^{str}-y^{str}_i) = \sigma_E \sqrt{g}\delta(y-y_i),
\label{eq:tensionrel}
\ee
where $\sigma_s$ and $\sigma_E$ denote the effective tensions in the string and Einstein frames, respectively. Note that the left hand side of \eqref{eq:tensionrel} is written in terms of the string frame metric ($g_{str}$) and radial coordinate $y^{str}$. The conversion between $\sigma_s$ and $\sigma_E$ results from the Weyl rescaling which relates both frames. 

For the class of objects consider here, the string frame tensions read $\sigma_s^{(1)}=\alpha_1$ and $\sigma_s^{(5)}=\alpha_5 e^{-D-3L}$, where the upper index denotes the number of spatial dimensions these objects wrap. Here $\alpha_1$ and $\alpha_5$ are numbers denoting the bare tensions, and will be determined from the boundary conditions at the incision radius. The factor of $e^{-D-3L}$ in $\sigma^{(5)}_s$ accounts for the inverse powers of the effective 6-dimensional string coupling. 

Putting everything together, we find 
\be
\sigma_E^{(1)} = \sigma_s^{(1)} e^{2D} \quad , \quad \sigma_E^{(5)}= \sigma_s^{(1)} e^{D-3L},
\label{eq:Etension}
\ee
so the overall effective tension at $y=y_i$ reads
\be
\sigma_{eff} = \alpha_1 e^{2D} +\alpha_5 e^{D-3L}.
\label{eq:efftension}
\ee

Assuming continuity of the metric and fields at $y=y_i$, a nontrivial tension such as the one in \eqref{eq:efftension} leads to a jump discontinuity for the solution. This can easily be seen by going to a frame in which the metric takes an FLRW-like form:
\be
ds^2 = dw^2 + \tilde a(w) \, d\gamma d\bar\gamma
\label{eq:FRW}
\ee
\begin{align}
\frac{1}{\tilde a}\left(\frac{d \tilde a}{dw}\Big|_{w\to w_i^-}-\frac{d \tilde a}{dw}\Big|_{w\to w_i^+}\right) &= \sigma_{eff}\label{eq:abc}\\
\frac{d D}{dw}\Big|_{w\to w_i^-}-\frac{d D}{dw}\Big|_{w\to w_i^+}&= -\partial_D\sigma_{eff}\label{eq:Dbc}\\ 
\frac{d L}{dw}\Big|_{w\to w_i^-}-\frac{d L}{dw}\Big|_{w\to w_i^+}&= -\frac13\partial_L\sigma_{eff}\label{eq:Lbc}
\end{align}
where $w_i$ denotes the position of the boundary in the frame \eqref{eq:FRW} and $w\to w_i^\pm$ means we approach the boundary from the exterior ($+$) and from the interior ($-$).

As advertised, we will now focus on a family of solutions to the bulk equations of motion derived from the action \eqref{eq:3daction} and to the boundary conditions \eqref{eq:abc}-\eqref{eq:Lbc} at $y=y_i$, which are free of singularities in the exterior region.

\subsection{Gluing to a linear dilaton background}

An exact solution of the boundary equations can be obtained by considering the situation in which the exterior geometry corresponds to a linear dilaton background at large radial positions. The transition between the interior and the exterior solutions is accomplished by the insertion of a thin shell at $y=y_i$ with tension given by \eqref{eq:efftension}. It is important to note that this ansatz implies a jump in the flux at $y=y_i$. Therefore, the exterior bulk equations of motion should come from a different potential. 

We parametrize this jump in the flux with an integer $\delta N_5$, foreshadowing the string-theoretic origin of these boundary conditions, which we will discuss in the next section. In terms of this parameter, the piecewise-defined potential reads
\be
{\cal V}=\begin{cases}
		 -6e^{2D-2L} +2Q_5^2e^{2D-6L}+2Q_1^2 e^{4D} , & y <  y_i	\\
		-6e^{2D-2L} +2(Q_5+\delta N_5)^2e^{2D-6L}+2(Q_1-\delta N_5)^2 e^{4D}, & y >  y_i
		\end{cases}
		\label{eq:pw-potential}
.\ee

A continuous solution for the bulk equations of motion derived from the above potential can be written in the following form
 \be
ds^2 = e^{-2D}\left(f dy^2+a d\gamma d\bar\gamma\right) 
\ee
\be
e^{-2D}= \begin{cases}
		  Q_1^2 Q_5\left(e^{2y}\left(\lambda -\frac{\delta N_5}{Q_1} e^{-2y_i}\right)+1\right)^2\left(1+\frac{\delta N_5}{Q_5}e^{2(y-y_i)}\right) & y <  y_i	\\
		  Q_1^2Q_5\left(\lambda e^{2y}+1-\frac{\delta N_5}{Q_1}\right)^2\left(1+\frac{\delta N_5}{Q_5} \right) & y >  y_i
		\end{cases}
\label{eq:glued-sol}
\ee
\be
a=\begin{cases}
		 \frac{e^2y}{e^{2y}\left(\lambda -\frac{\delta N_5}{Q_1} e^{-2y_i}\right)+1} & y<  y_i	\\
		\frac{e^{2y}}{\lambda e^{2y}+1-\frac{\delta N_5}{Q_1}} & y >  y_i
		\end{cases}
		\label{eq:gluea-sol}
\ee 
 \be
f=e^{2L} = \begin{cases}
		  Q_5+\delta N_5 e^{2(y-y_i)}, & y <  y_i	\\
		  Q_5+\delta N_5 , & y >  y_i
		\end{cases}
\label{eq:glueL-sol}
.\ee
Notice that these solutions are labeled by two parameters, namely the jump in flux $\delta N_5$ and the gluing position $y_i$. Here these parameters are free and may take any values that do not lead to problematic features, such as a naked singularity. We will fix their values at the end by matching to the background \eqref{eq:sol0} in the IR. 

Accounting for the form of the ansatz on either side of the gluing surface, the equations arising from the boundary conditions take the following form, where $\tilde a~=~e^{-2D} a$:
\begin{align}
\frac{e^{D}}{\sqrt{f}}\frac{1}{\tilde a}\left(\frac{d \tilde a}{dy}\Big|_{y\to y_i^-}-\frac{d \tilde a}{dy}\Big|_{y\to y_i^+}\right) - \sigma_{eff}\Big|_{y=y_i} &= 0 \label{eq:gabc}\\
\frac{e^{D}}{\sqrt{f}}\left(\frac{d D}{dy}\Big|_{y\to y_i^-}-\frac{d D}{dy}\Big|_{y\to y_i^+}\right)+\partial_D\sigma_{eff}\Big|_{y=y_i}&= 0\label{eq:gDbc}\\ 
\frac{e^{D}}{\sqrt{f}}\left(\frac{d L}{dy}\Big|_{y\to y_i^-}-\frac{d L}{dy}\Big|_{y\to y_i^+}\right)+\frac13\partial_L\sigma_{eff}\Big|_{y=y_i}&=0. \label{eq:gLbc}
\end{align}
The exact solution to these equations turns out to be as simple as one could have hoped, with
\be
\alpha_1 = -\delta N_5 \,\,\, , \,\,\, \alpha_5 = \delta N_5 
\label{eq:g-alpha}
. \ee
Notice that neither $\delta N_5$ nor $y_i$ are fixed by these equations. This is quite natural, as the only condition for the solution to make sense is that the shell satisfy the analog of Gauss' law. Some constraints between these variables will arise when we also ask the background to reproduce \eqref{eq:sol0} in the IR.  
 
Looking at $\alpha_1$, it seems that introducing a negative tension object is unavoidable. This is not necessarily important from the perspective of the effective theory, but certainly works against any interpretation in terms of standard string theory objects (recall that there is no gauged $\mathbb{Z}_2$ symmetry in our piecewise geometry, so orientifold planes are off the market). In section \ref{sec:10DPicture}, we will see that this pathology can be overcome for the case in which the compact manifold ${\cal M}_4$ is $K3$. 

In close relation to this observation, let us show here a curious aspect about the overall effective tension \eqref{eq:efftension} once evaluated in the solution \eqref{eq:g-alpha}. In particular, it is easy to check that the membranes become tensionless at a finite radius:
\be
\sigma_{eff}=0 \,\,\, {\rm for} \,\,\, e^{2y_i}=\frac{2\delta N_5+Q_5-Q_1}{\lambda Q_1}. \label{eq:3denhancon}
\ee         
This remarkable fact will take on a deeper meaning in the ten-dimensional story.

Finally, let us match the geometries above to \eqref{eq:sol0} in the IR. In order to achieve that, we need to impose
\be
\lambda -\delta N_5 e^{-2y_i} = -\lambda  
\ee
thus obtaining the following relation between our free parameters
\be
\delta N_5 = 2\lambda Q_1 e^{2y_i} \label{eq:3d-rel}
\ee
 Implicitly, there is a further constraint which arises from imposing regularity, {\it i.e.} $e^{2y_i}< e^{2y_s}$. Making use of \eqref{eq:3dsing}, this simply implies $\delta N_5< 2 Q_1$. It is instructive to check that this upper bound on $\delta N_5$ is enough to obtain an everywhere regular solution. We should focus on the potentially dangerous case of $Q_1<\delta N_5< 2Q_1$, which might induce a naked singularity in the exterior geometry, since $\lambda Q_5 e^{2y}+Q_1-\delta N_5$ vanishes for $e^{2y}=e^{2y^{ext}_s}$. It is simple to check that as long as $\delta N_5<2Q_1$, and assuming \eqref{eq:3d-rel}, then it is always true that $e^{2y^{ext}_s}<e^{2y_i}<e^{2y^{int}_s}$ (with $y^{int}_s$ given in \eqref{eq:3dsing}), thus avoiding any potential singularity.
Embedding this background into a 10-dimensional construction will lead to stronger constraints on $\delta N_5$, as detailed in section \ref{sec:10DPicture}.

Let us conclude by stating the background achieved after imposing \eqref{eq:3d-rel}:
\be
e^{-2D}= \begin{cases}
		  Q_1^2 Q_5\left(1-\lambda e^{2y}\right)^2\left(1+e^{2(y-y_{IR})}\right) & y <  y_i	\\
		  Q_1^2Q_5\left(\lambda e^{2y}+1-\frac{\delta N_5}{Q_1}\right)^2\left(1+\frac{\delta N_5}{Q_5} \right) & y >  y_i
		\end{cases}
\label{eq:fin-glued-sol}
\ee
\be
a=\begin{cases}
		 \frac{e^{2y}}{1-\lambda e^{2y}} & y<  y_i	\\
		\frac{e^{2y}}{\lambda e^{2y}+1-\frac{\delta N_5}{Q_1}} & y >  y_i
		\end{cases}
		\label{eq:fin-gluea-sol}
\ee 
 \be
f=e^{2L} = \begin{cases}
		  Q_5(1+e^{2(y-y_i)}), & y <  y_i	\\
		  Q_5+\delta N_5 , & y >  y_i
		\end{cases}
\label{eq:fin-glueL-sol}
\ee
where $\delta N_5$ is given by \eqref{eq:3d-rel} and we have defined 
\be
e^{2y_{IR}}=\frac{Q_5}{2\lambda Q_1}
,
\label{eq:yIR}
\ee
thus determining the scale below which the solution reproduces \eqref{eq:sol0}. 

Interestingly, the above system approaches a linear dilaton background of LST for large values of the radial coordinate. Thus for different radial slices, it appears to realize a non-trivial flow in the holographic dual which is driven by irrelevant deformations. Even if the flow initially resembles a $T\bar T$-deformed theory with the singular sign of the coupling, the pathologies of that flow are avoided by turning on a different set of irrelevant operators, which become dominant at intermediate energies. Finally, further up in the UV, the flow approaches the well-known trajectory triggered by $T\bar T$ with the opposite sign of the coupling, landing on a non-local field theory with Hagedorn growth.  

\section{Singularity resolution in ten dimensions} 
\label{sec:10DPicture}

As advertised, the solution derived above which interpolates between negative-$\lambda$ $T\bar T$-deformed $AdS_3$ and a linear dilaton background can be uplifted to 10-dimensional type IIB string theory. In this section we will demonstrate that, in this context, the singular behavior is naturally resolved by taking certain stringy effects into consideration. 

Let us first characterize the singularity itself by recalling the form of the 10-dimensional metric and dilaton corresponding to the extremal vacuum of the $F1$-$NS5$ system
\begin{align}
ds^2 &= f_1^{-1}\left(-dt^2+dx^2\right)+ f_5 \left(dr^2+r^2 d\Omega_3^2\right)+V^{1/2}ds_{{\cal M}_4}^2 \label{eq:10-metricF1NS5}\\
e^{2\Phi}&= g_s \frac{f_5}{f_1} \,\,\, , \,\,\, 
f_1=-1+\frac{r_1^2}{r^2} \,\,\, , \,\,\, f_5=1+\frac{r_5^2}{r^2}\label{eq:10-dilatonF1NS5}
\end{align}
with
\be
r_1^2= \frac{g_s^2\alpha' Q_1}{v}  \,\,\, , \,\,\, r_5^2=\alpha' Q_5
.\ee
Recall that $v$ is associated to the asymptotic volume of the Calabi-Yau ${\cal M}_4$ as $V=(2\pi)^4 \alpha'^2 v$.

As it stands, the above background is well defined only for $r<r_1$, as it features a naked singularity at $r=r_1$. The dilaton, and hence the effective string coupling, also diverges there. For $r>r_1$, CTC's develop, thus spoiling the causal structure of the exterior region. The actual singularity taking place at $r=r_1$ is of repulson type, which means that pointlike probe particles are repelled by it. We present the details of this identification in Appendix \ref{app:RepulsonDerivation}. The divergence of the string coupling near the singularity also means that one cannot reasonably trust \eqref{eq:10-metricF1NS5}, \eqref{eq:10-dilatonF1NS5}
in that region. However, the S-dual system is weakly coupled in that regime, and is thus more convenient to work with. In addition, we will find that the S-dual system provides a crucial handle which we can use to resolve the singularity.

\subsection{S-dual configuration and the enhan\c{c}on mechanism}

Under S-duality,\footnote{In the conventions adopted in this paper, S-duality maps $\Phi\to \tilde \Phi=-\Phi$ while leaving the Einstein frame metric invariant. Accordingly, we also take $g_s\to1/g_s$, $\alpha' \to g_s \alpha'$ and $v\to v/g_s^2$.}  \eqref{eq:10-metricF1NS5} and \eqref{eq:10-dilatonF1NS5} are mapped to the following background 
\begin{align}
    ds^2 &= Z_1^{-1/2}Z_5^{-1/2}(-dt^2+dx^2)+Z_1^{1/2}Z_5^{1/2}(dr^2+r^2 d\Omega_3^2)+Z_1^{1/2}Z_5^{-1/2} V^{1/2}ds^2_{{\cal M}_4} \label{eq:S10-metricF1NS5}
\end{align}\vspace{-0.5cm}
\begin{align}
    e^{2\tilde\Phi}&= g_s \frac{Z_1}{Z_5}\,\,\, ,  \,\,\, 
    Z_1=-1+\frac{\tilde r_1^2}{r^2} \,\,\, , \,\,\, Z_5=1+\frac{\tilde r_5^2}{r^2}\label{eq:S10-dilatonF1NS5}
\end{align}
with
\be
\tilde r_1^2= \frac{g_s\alpha' Q_1}{v}  \,\,\, , \,\,\, \tilde r_5^2=g_s \alpha' Q_5
.\ee
This clearly still has a naked singularity, occurring at $r=\tilde r_1$.
Remarkably, string theory has been shown to resolve such singularities on similar backgrounds, by means of the enhan\c{c}on mechanism \cite{JPP,Johnson:2001wm}. Quite importantly, this method hinges on the compact manifold ${\cal M}_4$ being $K3$, so we will consider this case in the rest of the analysis. Even though the physical setup differs from the one studied in the context of the enhan\c{c}on, many of its features bear a strong resemblance to the resolution we will propose for the background \eqref{eq:S10-metricF1NS5}-\eqref{eq:S10-dilatonF1NS5}. We will therefore review the key points about the original mechanism before discussing how it applies to the $T\bar{T}$-deformed background.  

\subsubsection{Brief review of the enhan\c{c}on mechanism}

The enhan\c{c}on mechanism was originally proposed in \cite{JPP} as a stringy resolution of repulson-type singularities in certain type II backgrounds associated to $D(p+4)$-branes wrapping a $K3$ manifold. It was principally motivated by the holographic realization of the Coulomb branch of ${\cal N}=2$ Super-Yang-Mills theory and the Seiberg-Witten description of its moduli space \cite{Seiberg:1994rs,OnlyPeet,Alberghi:2002tu,Benini:2008ir}. For $p=1$, it was subsequently extended to include arbitrary $D1$-brane fluxes \cite{Johnson:2001wm,Constable:2001fe}. Non-extremal solutions were also studied in this context \cite{Dimitriadis:2002xd,Dimitriadis:2003ya,Dimitriadis:2003ur}. In particular, the physics of the enhan\c{c}on was found to be vital for the second law of black hole thermodynamics to hold \cite{Johnson:2001us}.   
 
We will herein focus on the extremal background associated to $N_5$ $D5$-branes wrapped on $S^1\times K3$ together with $N_1$ $D1$-branes wrapping the $S^1$ and homogeneously smeared over the $K3$. When $D5$-branes wrap a $K3$ manifold, they acquire an induced negative $D1$ charge and tension~\cite{Bershadsky:1995qy,Green_1997,Dasgupta_1998,Bachas_1999,Giddings:2001yu} (see also Appendix \ref{app:PB}). This induced charge is crucial, as it brings about singularities. In our context, this phenomenon will also provide a way to justify the presence of the negative tension objects found in section~\ref{sec:3DPicture}. 

For concreteness, let us consider the case in which $N_5>N_1$, resulting in an overall negative $D1$ charge, so having
\begin{align}
     ds^2 &=\bar Z_1^{-1/2}\bar Z_5^{-1/2}(-dt^2+dx^2)+\bar Z_1^{1/2}\bar Z_5^{1/2}(dr^2+r^2 d\Omega_3^2)+\bar Z_1^{1/2}\bar Z_5^{-1/2} V^{1/2}ds^2_{{\cal M}_4} \label{eq:enhancon-metric}
\end{align}\vspace{-0.5cm}
\begin{align}
    e^{2\bar\Phi}&= g_s \frac{\bar Z_1}{\bar Z_5}\,\,\, ,  \,\,\, 
     \bar Z_1=1-\frac{\bar r_1^2}{r^2} \,\,\, , \,\,\, \bar Z_5=1+\frac{\bar r_5^2}{r^2}\label{eq:enhancon-dilaton}
\end{align}
with
\be
\bar r_1^2= \frac{g_s\alpha' \bar Q_1}{v}  \,\,\, , \,\,\, \bar r_5^2=g_s \alpha' \bar Q_5
\ee
and $\bar Q_1= N_5-N_1$, $\bar Q_5=N_5$. Again, $V=(2\pi)^4\alpha'^2 v$ denotes the asymptotic volume of the $K3$ and, moreover, we will take $v>1$. 

The repulson singularity occurs at $r=\bar r_1$, and for $r<r_1$ the geometry becomes ill-defined. Note that this behaviour is similar to that of the background \eqref{eq:S10-metricF1NS5}-\eqref{eq:S10-dilatonF1NS5}, but instead with the exterior geometry ($r>\bar r_1$) as the physically meaningful region. In this picture, however, the singularity is viewed as an artifact of ignoring an enhanced symmetry when the running $K3$ volume (in string frame) reaches the stringy scale $V^* = (2\pi)^4\alpha'^2$. More precisely,
\be
    V(r)=V \frac{\bar Z_1}{\bar Z_5} = V^* \,\,\, {\rm for} \,\,\, r=\bar r_e,
    \label{eq:enhancon-cond}
\ee
where $\bar r_e$ is the so-called ``enhan\c{c}on radius,'' which takes the form
\be
    \frac{\bar r_e^2}{g_s\alpha'} = \frac{\bar Q_5+\bar Q_1}{v-1}=\frac{2 N_5-N_1}{v-1}.
\ee
Note that the existence of the enhan\c{c}on radius is not necessarily tied to the presence of a singularity, as we could impose $N_5<N_1<2N_5$, which has $\bar r_e^2>0$. However, we will focus on the case in which its appearance is tied to the resolution of repulson singularities -- that is, $N_5>N_1$. Furthermore, note that in this case the enhan\c{c}on radius always sits outside of the singularity, {\it i.e.} $\bar r_e> \bar r_1$.

At the enhan\c{c}on radius, fivebrane probes  become tensionless and diverge in size, forming a shell which introduces new boundary conditions and excises the singular region $r< \bar r_e$. This can be seen by studying the dynamics of probe branes in the background \eqref{eq:enhancon-metric}-\eqref{eq:enhancon-dilaton}. For a supersymmetric $D5$-brane probe wrapping the $K3$, the probe tension is 
\begin{equation} \label{eqn:TEffEnhancon}
T_{eff}=T_1\left(\frac{T_5}{T_1} V \bar Z_1- \bar Z_5\right)=T_1(v \bar Z_1-\bar Z_5),
\end{equation} 
where $g_s T_1=(2\pi)^{-1}\alpha'^{-1}$ and $g_s T_5=(2\pi)^{-5} \alpha'^{-3}$ denote the bare tensions of $D1$- and $D5$-branes respectively.
\eqref{eqn:TEffEnhancon} vanishes precisely when the condition \eqref{eq:enhancon-cond} is met, {\it i.e.} at the enhan\c{c}on radius. As one pushes the probe closer towards the singularity, the tension becomes negative. Furthermore, there is no way to move the probe to $r<\bar r_e$ without breaking supersymmetry. On the other hand, a $D1$-brane probe is insensitive to any of these effects and is able to move freely towards the origin. Similar considerations also apply for composite objects made out of bound states between $D1$- and $D5$-branes, as long each fivebrane is dressed with at least one $D1$~\cite{Constable:2001fe}.

In the geometry of~\cite{JPP}, the authors argue that the delocalization of branes at $\Bar r_e$ signals a natural set of boundary conditions to impose. To excise the singular interior region $r<\bar r_e$, they replace it with a flat Minkowski spacetime, and posit that the fivebranes which source the geometry form a shell at $r=\bar r_e$. The same sewing procedure can be performed by replacing the interior with a (non-singular) extremal geometry, where only a fraction of the sources delocalize over a shell at $\bar r_e$~\cite{Johnson:2001wm}.  
The latter case is important, because we would like to resolve the singularity in \eqref{eq:S10-metricF1NS5}-\eqref{eq:S10-dilatonF1NS5} using a variant of the enhan\c{c}on mechanism for the interior region, but our geometries of interest are not flat.

As an illustration, here we consider a realization of the above mechanism, giving rise to a stitched solution which is well-behaved. We place all of the $D1$-brane sources at the origin, and we trade them for flux, whereas only some of the $D5$-branes, namely $N_5-\delta N_5$, are placed at the origin. The remaining $\delta N_5$ fivebranes sit at an incision radius $r_i$, and they source flux only for the exterior region ($r>r_i$). The metric takes the form \eqref{eq:enhancon-metric}-\eqref{eq:enhancon-dilaton}, with the harmonic functions now given by
\be
    \bar Z_1 = \left\lbrace\begin{array}{lcr}
            1-\frac{\bar r_{1+}^2-\bar r_{1-}^2}{ r_i^2}-\frac{\bar r_{1+}^2}{r^2} & , & r<r_i \\
            1-\frac{\bar r_{1+}^2}{r^2} & , & r>r_i
                \end{array} \right.
    \qquad 
            \begin{array}{ccl}
                \bar r_{1-}^2 & = & \bar{r}^2_{1+}-\frac{g_s\alpha' \delta N_5}{v}\\
                \bar r_{1+}^2 & = & \frac{g_s\alpha' \bar Q_1 }{v}
            \end{array}
\ee 
\be
\bar Z_5 = \left\lbrace\begin{array}{lcr}
1+\frac{\bar r_{1+}^2-\bar r_{5-}^2}{ r_i^2}+\frac{\bar r_{5-}^2}{r^2} & , & r<r_i \\
1+\frac{\bar r_{5+}^2}{r^2} & , & r>r_i
\end{array} \right.
\quad
\begin{array}{ccl}
\bar r_{5-}^2 & = & \bar{r}^2_{5+}-\frac{g_s\alpha' \delta N_5}{v}\\
\bar r_{5+}^2 & = & \frac{g_s\alpha' \bar Q_5 }{v}
\end{array}
.\ee 
As long as $\bar Q_1-\delta N_5=N_5-N_1-\delta N_5<0$, the above geometry is regular for any incision radius $r_i> \bar r_{1+}$.
This picture passes a number of tests.  In particular, the metric on moduli space as derived from the potential of a probe brane in the geometry
can be placed in clear correspondence with the related Coulomb branch metric in large-$N$ $SU(N)$ Seiberg-Witten theory~\cite{JPP,OnlyPeet}. 

The surface stress tensor associated to the junction at $r=r_i$ can be obtained from the Israel junction conditions \cite{Israel:1966rt}.
We absorb the $8\pi G$ prefactor into the definition of the surface stress tensor $S_{AB}$, as it does not play any role in the following. We have:
\be
    S_{AB}= \gamma_{AB}-G_{AB} \gamma^C_C  \,\, , \, \, \{A,B\} \in \{t,x,S^3, K3 \}
    \label{eq:surface-stress}
\ee
where
\be
    \gamma_{AB} = K^+_{AB}+K^-_{AB},
\ee
and
\be
    K_{AB}^{\pm} = \mp \frac{1}{2\sqrt{G_{rr}}}\partial_r G_{AB}, 
\ee
where $K^+_{AB}$ ($K^-_{AB}$) is the extrinsic curvature computed by approaching the shell from the exterior (interior) geometry, and $G_{AB}$ is the metric in the Einstein frame. For the case at hand, the surface stress tensor reads
 \begin{align}
S_{\mu\nu}&= \frac{1}{2\sqrt{G_{rr}}}\left(\frac{\Delta \bar Z_1'}{\bar Z_1}+\frac{\Delta \bar Z_5'}{\bar Z_5}\right) G_{\mu\nu} \,\, , & \{\mu,\nu\}\in \{t,x\} \\
S_{ij}&= 0 \,\, , & \{i,j\}\in S^3 \\
S_{ab}&= \frac{1}{2\sqrt{G_{rr}}}\left(\frac{\Delta \bar Z_5'}{\bar Z_5}\right) G_{ab} \,\, , & \{a,b\}\in K3
\end{align}
where we have defined $\Delta f' = f' \Big|_{r\to r_i^+}-f' \Big|_{r\to r_i^-}$.
Note the tension on the $S^3$ vanishes, as one expects for a BPS configuration.  Furthermore, the surface stress tensor is proportional to the probe brane tension \eqref{eqn:TEffEnhancon}:
\be
S_{\mu\nu} \sim -\delta N_5 (v \bar Z_1(r_i)-\bar Z_5(r_i)) G_{\mu\nu}=-\delta N_5 T_{eff} G_{\mu\nu}.
\ee
This is consistent with the claim that the shell is formed by $\delta N_5$ $D5$-branes. From the above expression we also conclude that the surface tension vanishes precisely at the enhan\c{c}on radius. Furthermore, the effective tension satisfies the Weak Energy Condition (WEC) only for $r_i\geq \bar r_e$. Many of these considerations will play an important role in the analysis to follow. 

\subsubsection{Glued solution and singularity resolution}

Having introduced the physics of the enhan\c{c}on, we are ready to resolve the singular behaviors of the S-dual background in \eqref{eq:S10-metricF1NS5}-\eqref{eq:S10-dilatonF1NS5}. Recall that the singularity occurs at $r=\tilde r_1$, where the function $Z_1$ vanishes. Let us recall the main differences between this setup and the one reviewed in previous subsection. For the geometry \eqref{eq:S10-metricF1NS5}-\eqref{eq:S10-dilatonF1NS5}, the IR region -- the interior -- is the well-defined region. We will search for a regular solution by stitching to a different background in the exterior, while requiring the interior region to reproduce the main features of \eqref{eq:S10-metricF1NS5}-\eqref{eq:S10-dilatonF1NS5} (in particular, the fluxes have to be given by the integers $Q_1$, $Q_5$). The difference between the interior and exterior fluxes will be parametrized by a single integer $\delta N_5$ and the incision will be made at a given radial position $r_i$.

In analogy with the construction described previously, the solution takes the form  \eqref{eq:S10-metricF1NS5}-\eqref{eq:S10-dilatonF1NS5} with the following piecewise-defined harmonic functions
	\begin{align}
		Z_1 &= \begin{cases}\label{eq:Z1glued}
		 1 - \frac{\tilde r_{1-}^2+\tilde r_{1+}^2}{r_i^2} + \frac{\tilde r_{1-}^2}{r^2}, & r < r_i \\
		 1 +\frac{\tilde r_{1+}^2}{r^2}, & r > r_i
		\end{cases}\\[1em] 
		Z_5 &= \begin{cases} \label{eq:Z5glued}
		 1 +\frac{\tilde r_{5+}^2-\tilde r_{5-}^2}{r_i^2}+\frac{\tilde r_{5-}^2}{r^2}, & r < r_i	\\
		 1 + \frac{\tilde r_{5+}^2}{r^2}, & r > r_i
		\end{cases}
	\end{align}	
with 
\begin{align}
\tilde r_{1-}^2= \frac{g_s \alpha' Q_1}{v} \qquad & , \qquad \tilde r_{1+}^2= \tilde r_{1-}^2 -\frac{g_s\alpha' \delta N_5}{v} \\[1em]
\tilde r_{5-}^2= g_s \alpha' Q_5 \qquad & , \qquad \tilde r_{5+}^2= \tilde r_{5-}^2 +g_s\alpha' \delta N_5
.\end{align}	
By construction, the metric is continuous at $r=r_i$. We now proceed to characterize the properties of the solution. Note that at this point the interior region does not look like the one we wanted to resolve. Below, we will impose some additional constraints on the parameters, namely $\delta N_5$ and $r_i$, in order to remedy this. 

Let us first find the general conditions under which the above solution is regular everywhere. It turns out that in order for $Z_1$ to not vanish at any radial position, we need the following relation to be hold:
\be
\frac{r_i^2}{g_s\alpha'}> \frac{\delta N_5-Q_1}{v}
.\ee
Note in particular that, in case of having $\delta N_5<Q_1$, the above inequality holds
for any choice of $r_i$. 

On the other hand, the exterior geometry has an enhan\c{c}on radius, at which the running volume of the $K3$ becomes of order the string scale, {\it i.e.} $V(r)=V^*$:
\be
\tilde r_e^2 = g_s \alpha' \frac{Q_5-Q_1+2\delta N_5}{v-1}.
\label{eq:enhancon}
\ee
As a further check that the above scale behaves as an enhan\c{c}on radius, we can compute the effective tension associated to a BPS $D5$-brane probe wrapped on $S^1\times K3$. The details of this calculation can be found in Appendix \ref{app:PB}, but the main result again looks like:
\be
T_{eff}= T_1 \left(v Z_1-Z_5\right)
\label{eq:D5efftension}
,\ee
so the effective tension again vanishes for $r=\tilde r_e$. Requiring a positive probe brane tension also induces a constraint on the incision radius $r_i$: in order to have well-defined probes on the geometry we should ask $r_i\geq\tilde r_e$. This condition is further supported by computing the surface stress tensor defined in \eqref{eq:surface-stress}. In particular, it vanishes along the $S^3$, due to the cancellation of inter-brane forces via supersymmetry. Moreover, along the $\{t,x\}$ directions, the tension computed from the surface stress tensor matches that of a shell of $\delta N_5$ $D5$ branes: 
\be
S_{\mu\nu} \sim -\delta N_5 (v Z_1-Z_5) G_{\mu\nu}=-\delta N_5 T_{eff} G_{\mu\nu}
\label{eq:S-surface-tension}
\ee
In order for these solutions to satisfy the WEC, we need $r_i\geq \tilde r_e$. 

We now proceed to impose one further condition on the background defined by the harmonic functions \eqref{eq:Z1glued} and \eqref{eq:Z5glued}, namely we want them to match \eqref{eq:S10-metricF1NS5}, \eqref{eq:S10-dilatonF1NS5} in the IR. This can be achieved by imposing:
\be
1-\frac{\tilde r_{1-}^2-\tilde r_{1+}^2}{r_i^2}=-1 \,\,\, \Rightarrow \,\,\, \delta N_5 = \frac{2 v r_i^2}{g_s\alpha'} 
\label{eq:match-cond}
.\ee
Plugging the above relation into \eqref{eq:enhancon} and imposing the WEC, we arrive at the following constraint:
\be
r_i^2 \leq r_{max}^2=g_s\alpha' \frac{Q_1-Q_5}{3v+1}
\label{eq:ri-max}
.\ee
Note in particular that this procedure is only possible for $Q_1>Q_5$. This is not a restrictive condition for our purposes; we are implicitly working in this regime, so that the corresponding $F1$-$NS5$ system will be weakly coupled in the IR. 

\begin{figure}[t]
  \centering
  \includegraphics[width=0.7\hsize]{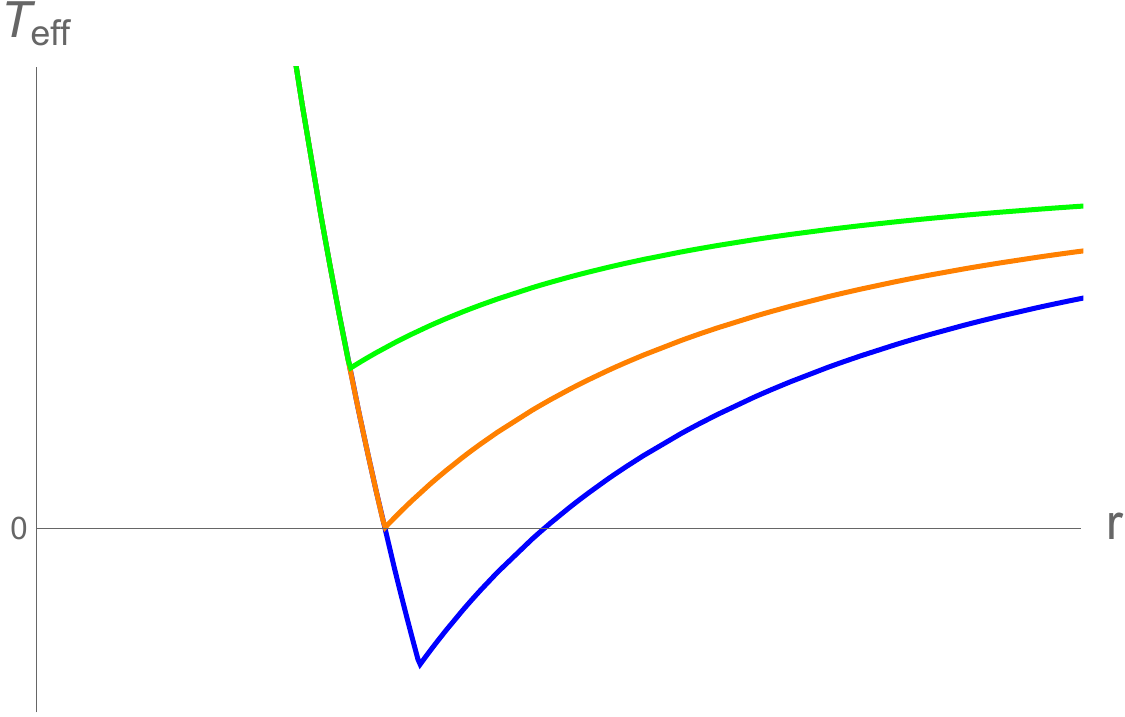}
  \caption{Effective tension as a function of the radius. The kink occurs at the corresponding junction position. Green, top: $r_i<r_{max}$ ($ r_e > r_i$). Orange, middle: $r_i=r_{max}$ ($r_e=r_i$). Blue, bottom: $r_i>r_{max}$ ($r_i< r_e$).}
  \label{fig:Teff}
\end{figure}

The moral of this story is the following: even though the WEC only demands $r_i\geq r_e$, it turns out that when we require the interior geometry to emulate \eqref{eq:S10-metricF1NS5}, the shell cannot be placed arbitrarily far away from the origin. The constraint arises from \eqref{eq:S10-dilatonF1NS5}. Importantly, the enhan\c{c}on radius coincides with the incision only when the bound \eqref{eq:ri-max} is saturated, {\it i.e.} $r_i=r_{max}$, so that
    \be
        r_e\Big|_{r_i=r_{max}} = r_i
    \ee
This is illustrated in figure~\ref{fig:Teff}. Note in particular that for $r_i>r_{max}$ the enhan\c{c}on lies at larger radius than the junction, and the surface tension \eqref{eq:S-surface-tension} at the interface becomes negative.  

So far we have shown that as long as $r_i\leq r_{max}$, we have $r_i\geq r_e$, and probe branes are well-behaved at any radial position. Moreover, the tension \eqref{eq:D5efftension} decreases when approaching the junction from either side, reaching a minimum at $r=r_i$ (see Fig. \ref{fig:Teff}). Although nothing dramatic occurs to the probes when they reach the junction if $r_i \neq r_e$, it is energetically favorable for them to stay there.

Among the possible $r_i\geq r_{max}$, it is natural to choose $r_i=r_{max}$, because that choice maximizes the region covered by the interior geometry. There is also a dynamical reason to take $r_i=r_{max}=r_e$, as only in this case do the branes at the junction become tensionless. Thus the configuration achieved for $r_i=r_{max}=r_e$ will have the least energy, and the branes at the junction will be uniformly distributed over the transverse $S^3$. 

\subsection{The $F1$-$NS5$ configuration}

Our long detour to the S-dual picture has led us to a regularized geometry given in terms of the piecewise-defined harmonic functions \eqref{eq:Z1glued}, \eqref{eq:Z5glued}. We can now S-dualize back and study the implications for our original singular background \eqref{eq:10-metricF1NS5}, \eqref{eq:10-dilatonF1NS5}. After the S-duality transformation, we obtain 
\begin{align}
    ds^2 &= f_1^{-1}\left(-dt^2+dx^2\right)+ f_5 \left(dr^2+r^2 d\Omega_3^2\right)+V^{1/2}ds_{{\cal M}_4}^2 \label{eq:10-glued-metricF1NS5}\\
    e^{2\Phi}&= g_s \frac{f_5}{f_1} \label{eq:10-glued-dilatonF1NS5}
\end{align}
with the harmonic functions given by
	\begin{align}
		f_1 &= \begin{cases}\label{eq:f1glued}
		 1 - \frac{r_{1-}^2+ r_{1+}^2}{r_i^2} + \frac{ r_{1-}^2}{r^2}, & r < r_i \\
		 1 +\frac{r_{1+}^2}{r^2}, & r > r_i
		\end{cases}\\[1em] 
		f_5 &= \begin{cases} \label{eq:f5glued}
		 1 +\frac{r_{5+}^2- r_{5-}^2}{r_i^2}+\frac{ r_{5-}^2}{r^2}, & r < r_i	\\
		 1 + \frac{ r_{5+}^2}{r^2}, & r > r_i
		\end{cases}
	\end{align}	
and with 
\begin{align}
 r_{1-}^2= \frac{g_s^2 \alpha' Q_1}{v} \qquad & , \qquad  r_{1+}^2=  r_{1-}^2 -\frac{g_s^2\alpha' \delta N_5}{v} \\
 r_{5-}^2=  \alpha' Q_5 \qquad & , \qquad  r_{5+}^2=  r_{5-}^2 +\alpha' \delta N_5
.\end{align}	
The relation \eqref{eq:match-cond}, maximal incision \eqref{eq:ri-max} and the enhan\c{c}on radius \eqref{eq:enhancon} are modified here by the appropriate powers of $g_s$. 

To check that the above solution corresponds to our glued solution in the effective 3-dimensional theory, we perform the LST decoupling limit by taking $g_s\to0$ after the change of variables $r^2=g_s^2\alpha' u^2$ (together with the redefinition $r_i^2=g_s^2\alpha' u_i^2$). We find:
\begin{align}
		f_1 &= \begin{cases}\label{eq:f1glued-dec}
		 1 - \frac{\delta N_5}{v u_i^2} + \frac{ Q_1}{v u^2}, & u < u_i \\
		 1 +\frac{Q_1-\delta N_5}{v u^2}, & u > u_i
		\end{cases}\\[1em] 
		g_s^2 f_5 &= \begin{cases} \label{eq:f5glued-dec}
		 \frac{\delta N_5}{u_i^2}+\frac{Q_5}{u^2} , & u < u_i	\\
		 \frac{Q_5}{u^2} \,\,\, , & u > u_i
		\end{cases}
	.\end{align}	
Let us pause here to comment on the region in which this background is weakly coupled, or equivalently, where an analysis via perturbative string theory is valid. Evaluating the dilaton field of the above, it can be easily seen that for $Q_1>Q_5$ the effective string coupling is small all the way up from the deep IR ($u\to 0$) to $u\sim u_i$. Near the junction, the coupling diverges and the picture becomes untrustworthy. It is thus in the S-dual system that the string coupling becomes weak in that region, and the junction acquires a natural interpretation as a shell of $D5$-branes. In the exterior geometry of the NS case, a perturbative description is valid for sufficiently large values of $u$, as the system approaches a linear dilaton background.	

Moving from this 10-dimensional picture down to the effective 3-dimensional theory amounts to following the steps listed in section \ref{sec:JJbarIntro}. After changing coordinates from $u$ to $y$ with $v u^2 = Q_1 \lambda e^{2y}$, and redefining $v u_i^2 = Q_1 \lambda e^{2y_i}$, we arrive at the solution described in section \ref{sec:3DPicture}. Finally, the relation \eqref{eq:3d-rel} is no more than the matching condition \eqref{eq:match-cond} after the S-duality transformation and the subsequent change of coordinates described above. 

With this 10-dimensional perspective, we see that the ability to find a 3-dimensional regular solution with a jump in flux parametrized by a single integer $\delta N_5$ is a manifestation of the fact that wrapping $D5$-branes on $K3$ generates an effective negative $D1$ tension and  charge~\cite{Bershadsky:1995qy,Green_1997,Dasgupta_1998,Bachas_1999,Giddings:2001yu}. Furthermore, the curious effect found in \eqref{eq:3denhancon} can now be understood as a consequence of the presence of an enhan\c{c}on scale in 10-dimensions. Thus, at least for the case where the compact manifold wrapped by the fivebranes is a $K3$, the singular behavior inherent to single-trace $T\bar T$-deformed backgrounds with negative $\lambda$ can be regularized by means of standard objects in type IIB string theory. The system maintains a strong resemblance to the well-understood physics of the enhan\c{c}on mechanism. 

Let us emphasize here that in this work we have achieved a particular UV completion, and other possibilities are by no means excluded. For instance, many of our considerations do not apply for the case when the Calabi-Yau 4-fold is a $T^4$, where a resolution might require involving unconventional string theory objects. We briefly comment on this situation in the next subsection.  

\subsection{$T^4$ and negative branes}

Before moving forward, we relate our setup to an exotic UV completion which has been suggested in the recent literature~\cite{Chakraborty:2020swe}. We will perform a nontrivial analytic continuation of this system to reveal the singular geometries obtained above. We herein take $\delta N_5 = 0$ so that there is no change in the behavior of $f_5$ at the junction, and we furthermore define:
	\begin{align}
		-1 \equiv 1+ \dfrac{r_{1+}^2 - r_{1-}^2}{r_i^2} = 1+ \dfrac{V^*}{V} g_s \alpha' \dfrac{\delta N_1}{r_i^2}
	.\end{align}
For purposes of demonstration, fix $r_i$ in this expression. Then, taking $\lambda \to -\lambda$ (which gives the constant $-1$ in $f_1$) is equivalent to analytically continuing $\delta N_1$ to a negative value. Such a procedure is subtle, because the analytic continuation in $\lambda$ can be achieved by a marginal current-current deformation of the string worldsheet, which should not change the supersymmetries preserved by the background~\cite{Petropoulos:2009jz}.\footnote{Specifically, a TsT deformation of the $F1$-$NS5$ system can be used to parametrize the $J\bar{J}$ deformation~\cite{Apolo:2019zai}. This makes the analytic continuation manifest: $f_5$ is left unchanged, whereas $f_1 \to f_1 + 2g_s^{-1}\lambda_{TsT}$. In that picture, $\lambda_{TsT} \mapsto -\lambda_{TsT}$ is equivalent to $\delta N_1 \mapsto -\delta N_1$.} Thus the analytic continuation in $\delta N_1$ must be performed whilst fixing supersymmetries, i.e. the branes at the junction cannot possibly be anti-branes. Instead, they are objects with negative tension and negative charge which are BPS with the D-branes in the geometry. 

This picture is analogous to the procedure used to derive the original enhan\c{c}on geometry~\cite{JPP}, where the authors extend the exterior geometry to all $r$ and analytically continue $Q_{1+}$ to $-Q_{5+}$. They fix the same unbroken supersymmetries as in the $Q_{1+}>0$ case. In that context, it was by switching from a $T^4$ to a $K3$ compactification that the authors were able to avoid introducing an object with negative effective tension and recover the shell of branes. Due to the negative induced $D1$-brane charge that $D5$-branes carry when wrapped on $K3$~\cite{Bershadsky_1996,Green_1997,Dasgupta_1998,Bachas_1999}, these geometries can be generated from wrapped $D5$-branes alone. However, it is perfectly valid to insist on keeping the $T^4$.

To understand this geometry fully, we must rely on the negative-tension objects which are prescribed to us by the string theory literature. An option which was well-studied in~\cite{Chakraborty:2020swe} made use of so-called ``negative branes,'' which have SU$(0|M)$ gauge symmetry groups, {\it i.e.} their Chan-Paton factors are Grassmann numbers~\cite{dijkgraaf2016negative}. This option is certainly available for either choice of compactification manifold, but such objects tend to introduce exotic physics, such as signature changes and closed timelike curves. Notably, this option was also available in the original enhan\c{c}on backgrounds. In that context, however, the restriction to the more standard taxonomy of extended objects in string theory uncovered elegant physics, and we therefore chose to restrict ourselves to these in this work. 

There is one negative-tension, negative-charge candidate available to us in this scenario: orientifold planes, or $O1_-$-planes. These objects have no motion collective coordinates. They are nondynamical, because they introduce boundary conditions on the spacetime and fluxes which can be viewed (in a simplified sense) as a $\mathbb{Z}_2$ quotient of two copies of the system mirrored across the fixed plane. Orientifolds have been considered as an extension of the original enhan\c{c}on results, in~\cite{Jarv:2000zv}. We have restricted the scope of this initial work to a resolution via branes, but we find the prospect of a compactification via $O$-planes tantalizing, and we hope to address it in subsequent work.

\section{Comments on the spectrum of excited states} \label{sec:comments}

In this last section, we briefly expand on some ideas which may provide a more complete insight about the nature of the regular backgrounds studied so far. 

In particular, the glued geometry constructed in sections~\ref{sec:3DPicture} and~\ref{sec:10DPicture} corresponds to the deformation of the RR vacuum state, and thus looks like the pure $AdS_3$ vacuum at small enough radial positions. The spectrum of low energy excitations above the undeformed ($AdS_3$) vacuum can be obtained by studying the dynamics of long strings on top of it. The latter amounts to considering the (spectrally flowed) continuous representations of $SL(2,\mathbb{R})$~\cite{Maldacena:2000hw}. The excitations of the untwisted sector can be thought of as corresponding to the dynamics of a single block in the dual orbifold CFT. The spectrum of such long string states has been computed exactly in the single-trace $T\bar T$-deformed vacuum using standard techniques in~\cite{Apolo:2019zai}. This is possible because the deformed worldsheet sigma model remains solvable for any value of the deformation parameter.   
Unfortunately, in our setting, although the calculation remains the same in a portion of the interior geometry, the backgrounds presented herein do not correspond to the conventional single-trace $T\bar T$ deformation. Rather, they correspond to a more complicated trajectory in the space of QFT's, which is still unknown to us. The absence of singularities in our backgrounds is a strong indication of the avoidance of complex energy levels above a certain cutoff scale. 

We can nevertheless characterize the dominant contributions of this flow in different energy regimes. At low energies, we can argue that the spectrum of long strings stretched at radial distances lying well below the scale $y_{IR}$ introduced in section \ref{sec:3DPicture} must coincide with the $T\bar T$-deformed spectrum with negative coupling in a single block of the deformed orbifold QFT. This statement comes naturally from the fact that the sigma model is still solvable in this region, as explained above. Moreover, by means of expression \eqref{eq:yIR}, the region of analytic control can in principle be enlarged at will by considering a certain double-scaled limit in which $Q_1\to \infty$, $\lambda\to 0$ with $\lambda Q_1 \ll Q_5$. 

In the other extreme regime, we might consider very massive states which, as usual, are not described by perturbative dynamics, but by black holes. Obtaining consistent black hole states for these sort of glued constructions is a challenging task in the context of the enhan\c{c}on mechanism, and was studied in~\cite{Johnson:2001wm,Dimitriadis:2003ur,Dimitriadis:2002xd,Dimitriadis:2003ya}. This difficulty can in principle be attributed to a number of ambiguities which arise when determining the actual geometry associated to these states. One of the most severe ambiguities resides in the presence of two horizons, respectively for the interior and exterior geometries. In principle, there is no further insight from the supergravity equations which allows us to fix a particular relation between these parameters, which is an obstruction to determining their thermodynamics. Quite importantly, a branch of these solutions has been shown to violate the Weak Energy Condition in a broad regime of parameter space, being therefore characterized as unphysical~\cite{Dimitriadis:2003ur,Dimitriadis:2003ya}. 

Thus a generic treatment of black hole solutions over these backgrounds is beyond the scope of this work. It might be argued that the knowledge of the dynamics of the holographic IR CFT may provide some insight to overcome these issues, and that would be an interesting idea to pursue in future work. For now, we will instead focus on the particular cases which are free of such ambiguities: namely heavy states lying in the upper region of the mass spectrum. More concretely, solutions featuring a large horizon radius are simple in the sense that any subtlety related to the interior geometry or the details of the junction can be disregarded once the interior region lies completely within the horizon. The thermodynamics associated to these configurations therefore only care about the exterior geometry. The validity of these solutions is guaranteed as long as they do not develop an enhan\c{c}on radius outside the horizon, as we will comment below.

For concreteness, we will consider non-rotating charged black hole solutions. The charges correspond to the ones present on the exterior region, namely $Q_1^{ext}= Q_1-\delta N_5\equiv\tilde Q_1$, $Q_5^{ext}= Q_5+\delta N_5\equiv\tilde Q_5$. The string frame metric reads
\be
ds^2= h_1^{-1}(-K dt^2+dx^2)+h_5(K^{-1}dr^2+r^2 d\Omega_3^2)+ V^{1/2}ds^2_{K3}
\label{eq:BHext}
\ee
where
\be
h_1 = 1+\frac{\hat r_{1}^2}{r^2} \,\,\, , \,\,\, h_5 = 1+\frac{\hat r_{5}^2}{r^2}
\,\,\, , \,\,\, 
 K = 1-\frac{ r_0^2}{r^2} 
\label{eq:BHsolution}
\ee
with $\hat r_1$ and $\hat r_5$ given by
\begin{align}
\hat r_1^2 =  r_0^2 \sinh^2\alpha_1 \,\,\, &, \,\,\, \hat r_5^2 = r_0^2 \sinh^2\alpha_5  \nonumber \\
\sinh{2\alpha_1}=\frac{2g_s^2\alpha' \tilde Q_1}{vr_0^2} \,\,\, &, \,\,\, \sinh{2\alpha_5}=\frac{2\alpha' \tilde Q_5}{r_0^2}
.
\label{eq:BHparameters}
\end{align}
The dilaton and 3-form fluxes depend on the harmonic functions in the same way as for the extremal solutions, so we do not write them here.

The thermodynamics associated to these configurations has recently  been studied in \cite{Chakraborty:2020swe}, as they are associated to very massive states in positive coupling single trace $T\bar T$. As our geometries have the same asymptotic behaviour in the UV region, the conclusions of \cite{Chakraborty:2020swe} carry over to those same high-energy states here. As the relevant computations have been already performed in \cite{Chakraborty:2020swe}, we limit ourselves to report the relevant results here. To keep our discussion self-contained, we show some details of the calculation in Appendix \ref{app:BHentropy}. 

We will work in the decoupling limit, for which $\alpha_5\to\infty$ while $\alpha_1$ is kept fixed.
The entropy as a function of the dimensionless energy above extremality ${\cal E}$ (see Appendix \ref{app:BHentropy}) is accounted by the area of the compact horizon at $r=r_0$ and reads
\be
S = 2\pi \sqrt{\tilde Q_5} \sqrt{2 \tilde Q_1 {\cal E}+\lambda {\cal E}}
\label{eq:entropy}
\ee
hence making manifest that the dynamics of high energy states interpolates between the usual Cardy regime and the Hagedorn growth, with the entropy reaching its maximum at the critical energy
\be
{\cal E}_c = \frac{\tilde Q_1}{\lambda} 
.\ee 
For energies above this scale, the effective temperature becomes negative and the theory becomes highly non-local. We emphasize that none of the results stated above are new. We reproduce them here to illustrate the dynamics associated to high energy states in our setup, and to connect the resolved geometries presented herein to previous work. 

It is instructive to compute the minimal energy for which these considerations are valid. This amounts to computing the particular value of ${\cal E}$ for which the horizon $r_0$ meets an enhan\c{c}on radius. To find the enhan\c{c}on radius in this context, we have to go again to the S-dual geometry. Notice that the probe brane analysis is no longer insightful, as there is a nontrivial potential for the brane motion, due to the absence of supersymmetry. However, one can still define the enhan\c{c}on as the radial position at which the $K3$ becomes of the string scale, so obtaining
\be
\hat r^2_e = \frac{\hat r_{5}^2-v\hat r_{1}^2}{v-1}
\ee  
where, in an abuse of notation, we are denoting by $\hat r_{1}$, $\hat r_{5}$ to the corresponding S-dual quantities. By S-dualizing back, we get
\be
\hat r^2_e = \frac{1}{v-g_s^2}\left(g_s^2\hat r_{5}^2-v\hat r_{1}^2\right)
\ee  
and, taking the decoupling limit and expressing all quantities in terms of the energy above extremality, we finally obtain
\be
\frac{\hat r_e^2}{r_0^2} = \frac{\Tilde Q_5(\tilde Q_1+\lambda {\cal E})-\tilde Q_1^2}{\lambda {\cal E}(2\tilde Q_1+\lambda {\cal E})}
.\ee
From this, we derive the minimal energy condition as:
\be
\frac{\hat r_e^2}{r_0^2}<1 \,\,\, \Rightarrow \,\,\, {\cal E}>{\cal E}_{min} 
\ee
with
\be
{\cal E}_{min} = \frac{\tilde Q_5-\tilde Q_1}{\lambda}=\frac{ Q_5- Q_1+2\delta N_5}{\lambda} \label{eq:min-en}
\ee
The above quantity hence represents the minimal energy for which the large horizon analysis presented so far holds. To understand what happens at scales below \eqref{eq:min-en}, one needs a consistent glued black hole solution, featuring interior and exterior horizons and merging at a given radial position. As explained previously, those solutions are subtle, so we leave a more thorough analysis of this region of the spectrum to future work.

\section{Discussion}
\label{sec:Discussion}

In this work, we constructed a family of regular solutions of type IIB supergravity with Neveu-Schwartz flux which resolve the naked singularity in single-trace $T\bar T$-deformed backgrounds. The resolution procedure has some direct connections with the enhan\c{c}on mechanism, previously considered in the literature to resolve repulson-type singularities \cite{JPP,Johnson:2001wm}. Even though we reproduce the single-trace $T\bar T$-deformed background in the IR region, the solutions considered here do not correspond holographically to the same flow in the space of QFT's. In particular, the avoidance of curvature and dilaton singularities, as well as the absence of closed timelike curves, signal that this flow can be traced all the way to the UV region without developing complex energy levels. In the UV, the theory becomes non-local and develops Hagedorn scaling in the density of states. This is clear if one observes that the supergravity background reduces to a linear dilaton vacuum of Little String Theory in the asymptotic region. 

It is important to comment on the indirect nature of the mechanism we have proposed. In particular, because the effective string coupling in the NS-NS case grows strong near the singularity, a trustworthy description is only achieved by S-dualizing to a background featuring R-R fluxes. It is in this S-dual system that the physics of D-branes wrapped on $K3$ allows for a resolution via the enhan\c{c}on construction. This is the only consistent mechanism we have identified in which the singularity is resolved by standard dynamics in string theory; that is, without appealing to any exotic soliton to account for negative tensions. Let us nevertheless emphasize that a more direct resolution concerning the dynamics of fundamental strings and $NS5$-branes would be very interesting to achieve. The highly non-trivial action functional of $NS5$-brane probes makes this task computationally challenging.

Another exciting and in principle affordable possibility is to find a consistent truncation of the radial direction within a warped compactification \cite{Randall:1999ee, Randall:1999vf}. Achieving this may establish a non-trivial connection between single-trace $T\bar T$ and the finite cut-off holographic realization of double-trace $T\bar T$ \cite{MMV}. Accordingly, it may also be more natural to construct a ``single-trace'' analogy of the ``$T\bar T \, + \, \Lambda_2$'' flows recently introduced in \cite{Gorbenko:2018oov}, interpolating between globally AdS and dS spacetimes, given that dS is already compact in the spatial directions. We hope to address this point in the near future.    

Finally, within the context of the combined flows studied in this paper, there are two important gaps to be filled. One is to perform a more systematic study of black hole solutions joined at a finite radial position, in order to describe the dynamics of intermediate energy states. The other is to obtain these geometries from the worldsheet perspective---that is, to find a particular (and hopefully solvable) deformation of the string sigma model featuring these geometries as solutions of the field equations.\footnote{ Studying the possible deformations in terms of a gauged sigma model, as recently shown in \cite{Chakraborty:2020yka}, may be useful to achieve this task.}

\section*{Acknowledgements}
We are grateful to Eva Silverstein for proposing the key ideas developed in this project, and for her guidance as the results unfolded. We thank Gonzalo Torroba, Wei Song, Stephane Detournay, Victor Gorbenko, G. Bruno de Luca, Luis Apolo, Aitor Lewkowycz, and Alex Musatov for their input and insight on our findings. The work of LA was supported by the Knut and Alice Wallenberg foundation.
The work of EAC is supported by the US NSF Graduate Research Fellowship under Grant DGE-1656518. 
This research was partially completed at the Kavli Institute for Theoretical Physics, to which we are very grateful; as such, our collaboration was supported by the National Science Foundation under Grant No. NSF PHY-1748958.

\appendix
\section{Derivation of repulson behavior}
\label{app:RepulsonDerivation}

Technically, for the enhan\c{c}on mechanism to activate, we only need the running $K3$ volume to reach $V^* = (4\pi^2\alpha')^2$ at some locus in the string frame geometry. As far as the initial enhan\c{c}on results~\cite{JPP} are concerned, this behavior defines a repulson-type singularity. However, such a characterization raises concern for the interpretation of the singularity in the NS-NS sector, where the running $K3$ volume is constant. Here, we will confirm the repulson-type behavior by studying the original definition, which depends only on characteristics of the Einstein frame metric and is therefore unaffected by an S-duality transformation. The first works on repulsons~\cite{Repulson1,Repulson2,Repulson3} characterized their backgrounds of interest by studying the classical behavior of matter moving toward the singularity from the well-behaved region. As described in detail in~\cite{Repulson2}, the effective gravitational potential grows positive and becomes infinite as a massive test particle approaches the singularity. The particle is inevitably repelled, hence the name ``repulson.'' These analyses (c.f. section 5 of~\cite{Repulson2}) closely followed a textbook derivation from section 98 of~\cite{landau1975classical}. We will go through it here, both for the geometry in the original enhan\c{c}on results, as well as for our system of interest. 

In Einstein frame, the generic $D1$-$D5$ background (equivalently, $F1$-$NS5$ background) is rotationally invariant and has a timelike Killing vector, so the energy $E$ and momenta $P_i$ of the particle will be conserved. For simplicity, we will neglect momenta along $K3$ and $S^3$, and instead allow only for angular momentum $L$ around $x$. The equation of state for the system is
    \begin{align}
        S = - E t + L x + S_r(r), \qquad\qquad g^{\mu\nu} \dfrac{\partial S}{\partial x^\mu}\dfrac{\partial S}{\partial x^\nu} = m^2.
    \end{align}
This presents a differential equation for $S_r$, which we can solve with
    \begin{align}
        S_r(r) = \pm\int_{r_0}^r dr' \sqrt{-\dfrac{g_{rr}}{g_{tt}g_{x x }}\left(E^2 g_{x x} +m^2 g_{tt} g_{x x} + L^2 g_{tt}\right)} .
    \end{align}
Plugging this into $S$, we can now impose the Hamilton-Jacobi equation $\frac{\partial S}{\partial E} = 0$. This gives us a relation between $r$ and $t$:
    \begin{align} \label{eqn:tFromIntRRepulson}
        t = \int^{r}_{r_0} dr'\, \frac{E r'}{\sqrt{-g_{tt} \left( E^2 g_{rr} + m^2 g_{tt} g_{rr} + L^2 \frac{g_{tt}}{g_{x x}} g_{rr}\right)}},
    \end{align}
where the sign is fixed in this case by specifying that the particle starts from some radius $r_0$ far away from the singularity and falls in toward it. This is equivalent to (46) in~\cite{Repulson2}, where in that case $\frac{g_{rr}}{g_{x x}} = \frac{1}{r^2}$ and they take $g_{tt} = g_{xx}$ (i.e. they drop the sign difference). The simple way to view the repulson behavior in this expression is to note for what values of $r$ the denominator is pure complex. Generically, this will occur at some radius before the particle hits the singularity, irrespective of energy or angular momentum. Rather than a point of no return (an event horizon), there is a point of no advance, which the authors of~\cite{Repulson2} referred to as ``antigravity.''

In 10 dimensions, the Einstein frame metric has 
    \begin{align}
        g_{rr} = f_1^{1/4} f_5^{3/4}, \qquad g_{xx} = -g_{tt} = f_1^{-3/4}f_5^{-1/4}, 
    \end{align}
and the integrand takes the form
    \begin{align}
        \dfrac{E f_1^{5/8} f_5^{7/8}}{\sqrt{(E^2-L^2) f_1^{1/4} f_5^{3/4} - m^2 \sqrt{\frac{f_5}{f_1}}}}.
    \end{align}
As we move toward the singularity, $f_1$ shrinks to $0$ whereas $f_5$ does not. For $m\neq 0$ the argument to the square root in the denominator eventually becomes negative, irrespective of the energy $E$. Thus the repulson behavior is visible both for the original enhan\c{c}on geometry with $f_i=\bar Z_i$ in~\eqref{eq:enhancon-dilaton} as well as the deformed geometry with $f_i$ in~\eqref{eq:10-dilatonF1NS5}. The exact radii where the integrands become complex are attainable for both systems, although they have complicated and unenlightening forms. 

The massless case requires special treatment. We write the proper velocity $U$ as  
\begin{equation}
U_\mu = 
    \begin{pmatrix}
    E \\ L \\ U_r
    \end{pmatrix}.
\end{equation}
Enforcing $g_{\mu \nu } U^{\mu } U^{\nu }=0$ allows us to solve for $U_r$ in terms of $E, L$ and $r$. The proper velocity and acceleration, with affine parameter $\xi$, is
    \begin{align}
        U^r &= \dfrac{\partial r}{\partial \xi} = \pm\sqrt{E^2-L^2} \left(\dfrac{f_1}{f_5}\right)^{1/4} \\
        \partial_\xi U^r &= \mp(E^2-L^2)\dfrac{r_1^2 + r_5^2}{2r^3 f_1^{1/2} f_5^{3/2}}
    \end{align}
It is clear that the maximum value of $r(\xi)$ is the radius where $f_1$ vanishes, and that the singularity repels the geodesics with diverging strength as they approach the singularity.

\subsection*{A subtlety in the dimensional reduction}

There is a slight subtlety in the definition of repulson used above, as the Weyl factor used to derive the Einstein frame metric depends on the number of dimensions we choose to compactify. To illustrate this nuance, let us consider the system in only the $t,r,x$ coordinates and compactify the $S^3 \times {\cal M}_4$. Then:
    \begin{align}
        g_{rr} = f_1^{-3/4} f_5^{5/4}, \qquad g_{xx} = -g_{tt} = f_1^{-5/4}f_5^{3/4},
    \end{align}
and one finds that the integrand in~\eqref{eqn:tFromIntRRepulson} instead becomes 
    \begin{align} 
        \dfrac{E (f_1 f_5)^{7/8}}{\sqrt{(E^2-L^2)(f_1 f_5)^{5/4} -m^2 f_5^2 }}
    \end{align}
Massive particles will still not be able to reach the singularity in finite time. However, we argue that the most sensible definition of repulson behavior is the one which makes use of the 10-dimensional geometry, i.e. without any compactification. This convention minimizes ambiguity, although it is different from the one in~\cite{Repulson2}.

\section{Probe brane computation} \label{app:PB}

We are interested in a $D5$-brane probe on a background of the form
\begin{align}
    ds^2 &= Z_1^{-1/2}Z_5^{-1/2}(-dt^2+dx^2)+Z_1^{1/2}Z_5^{1/2}(dr^2+r^2 d\Omega_3^2)+Z_1^{1/2}Z_5^{-1/2} V^{1/2}ds^2_{{\cal M}_4}
\end{align}
\begin{align}
    e^{2\tilde\Phi}&= g_s \frac{Z_1}{Z_5}\,\,\, ,  \,\,\, 
    C_2= Z_1^{-1} dt\wedge dx \,\,\, , \,\,\, C_6=Z_5^{-1} dt\wedge dx  \wedge V \epsilon_{K3}
\end{align}
where $\epsilon_{K3}$ is the volume form on the unit $K3$.

The probe will be wrapped over the $K3$ and extended on the non-compact $t$ and $x$ directions. By fixing the static gauge, we identify the world volume coordinates 
\be
\{\zeta^a\} = \{ t,x,K3 \}
\ee
Moreover, we will only consider time dependence in the transverse directions. 

For the induced metric on the world volume we obtain
\begin{align}
g_{tt} &= -Z_1^{-1/2}Z_5^{-1/2} +Z_1^{1/2}Z_5^{1/2} v_T^2 \\
g_{x x} &= Z_1^{-1/2}Z_5^{-1/2}\\
g_{ij}&= V^{1/2}Z_1^{1/2}Z_5^{-1/2}g^{K3}_{ij} \,\,\, , \,\,\, \{i,j \in K3\}
\end{align}
where we have defined the transverse velocity $v_T^2= \dot{r}^2+r^2 (\dot{\theta}^2+\sin^2\theta \dot\varphi_1^2+\sin^2\theta \sin^2\varphi_1 \dot\varphi_2^2) $ and the angles $\{ \theta, \varphi_1, \varphi_2 \}$ are coordinates on the 3-sphere. Therefore, in a slow-moving approximation, we have
\be
\sqrt{-\det g} = \frac{V}{Z_5}\sqrt{\frac{Z_1}{Z_5}}\sqrt{1-Z_1 Z_5 v_T^2}\sqrt{\det g^{K3}}
\approx \frac{V}{Z_5}\sqrt{\frac{ Z_1}{Z_5}}\sqrt{\det g^{K3}}\left(1-\frac{Z_1 Z_5 v_T^2}{2}\right)\label{indexp}
.\ee
With the above expression in hand, we can evaluate the DBI action (note that we are setting the internal $U(1)$ gauge field to zero):\footnote{Recall that, according to the conventions followed in this paper, the bare tensions for the probe $D1$- and $D5$-branes are $g_s T_1=(2\pi)^{-1}\alpha'^{-1}$ and $g_s T_5=(2\pi)^{-5} \alpha'^{-3}$ respectively.}
\be
S_{DBI} = - T_5 \int d^6 \zeta \, e^{-\Phi} \sqrt{-\det g} 
 \approx -T_5 \int dt \, \frac{V}{Z_5} \left(1-\frac{Z_1 Z_5 v^2}{2}\right)
.\ee 
Now we proceed to evaluate the WZW coupling of the $D5$-brane to the 6-form flux. The latter has no transverse components, so the pullback is trivial, giving 
\be
S_{WZW} = T_5 \int d^6 \zeta \, C_{6}= T_5\int dt \, \frac{V}{Z_5}.
\ee 
However, here certain stringy effects must be taken into account. As explained in~\cite{Bershadsky:1995qy} (and reviewed in first section of~\cite{Giddings:2001yu}), by anomaly inflow there is an additional WZW coupling when wrapping a $D5$-brane on $K3$. This additional contribution is proportional to $\alpha'^2$ and roughly takes the form 
\be
\delta S_{WZW} \sim T_5 \alpha'^2 \int_{WV} C_{2}\wedge p_1(K3),
\ee 
where $p_1$ denotes the first Pontryagin class of $K3$. Given that $C_{2}$ does not have any component on $K3$, and since $p_1$ is quantized to a negative integer, it turns out that
\be
\delta S_{WZW} =- T_1 \int dt \, Z_1^{-1}. 
\ee 
Note the absence of a factor of $V$, given that $p_1$ is a topological invariant and, as such, is independent of the overall volume of $K3$. Here we have used that $T_1/T_5 =(2\pi)^4 \alpha'^2$.

Since the configuration is still BPS, the tension has to acquire a compensating term of the same form. By the above reasoning, this new term does not have any factor of $V$ either. Thus the relevant induced metric is the one corresponding to the subspace spanned by $t,x$, giving 
\be
\delta S_{DBI} = T_1 \int d^2\zeta e^{-\Phi}\sqrt{-\det g^{(t,x)}} \approx T_1 \int dt \frac{1}{Z_1} \left(1-\frac{Z_1 Z_5 v_T^2}{2}\right)  
.\ee
Putting all together we get
\begin{align}
{\cal L}_{probe} &\approx  \left(-T_5 V Z_5^{-1}+ T_1  Z_1^{-1}\right) \left(1-\frac{Z_1 Z_5 v_T^2}{2}\right)+ T_5 V Z_5^{-1}- T_1   \tilde{Z}_1^{-1}\\
&= \frac12 \left(T_5 V  Z_1- T_1 Z_5\right) v_T^2
,\end{align}
hence the effective tension reads
\be
T_{eff}=T_1\left(\frac{T_5}{T_1} V Z_1- Z_5\right),
\ee
as quoted in the main text.

\section{Computation of the black hole entropy} \label{app:BHentropy}

To keep the presentation self-contained, we review the basics of the computation done in \cite{Chakraborty:2020swe} which leads to the entropy \eqref{eq:entropy}. As explained in section \ref{sec:comments}, for sufficiently large mass, the relevant black hole solutions do not feature an enhan\c{c}on occurring outside the horizon, hence only the exterior geometry matters for describing their thermodynamics. 

It is natural to dimensionally reduce to five dimensions,\footnote{Of course, the same result can be obtained by reducing to three dimensions by following the steps depicted in section \ref{sec:3DPicture}.} where the Einstein frame metric reads
\be
ds^2 = -(h_1 h_5)^{-2/3}K dt^2 +(h_1 h_5)^{1/3}(K^{-1}dr^2+r^2d\Omega_3),
\label{eq:5d-metric}
\ee
with the harmonic functions $f_1$, $f_5$ and emblackening factor $K$ given in \eqref{eq:BHsolution}, \eqref{eq:BHparameters}. 

The energy associated to these configurations is accounted for by the ADM mass. The latter can be read off from the asymptotic structure of the metric:
\be
g_{tt}= -1 + \frac{8 G_N^{5d}}{3\pi} \frac{M}{r^2} + {\cal O}(r^{-4}) \, ,
\ee
which in this case yields 
\be
M = \frac{3\pi}{8 G_N^{5d}}\frac23 (\hat r_1^2+\hat r_5^2+\frac32 r_0^2) 
= \frac{vRr_0^2}{2g_s^2\alpha'^2}(\cosh2\alpha_1+\cosh{2\alpha_5}+1) \, ,
\ee
and we have used the five-dimensional Newton's constant of the form
\be
G_{N}^{5d}= \frac{G_N^{(10)}}{V (2\pi R)} = \frac{\pi g_s^2 \alpha'^2}{4vR}
.\ee  
The extremal mass is readily obtained by taking $r_0\to 0$, giving 
\be
M^{ext}=\frac{vR}{g_s^2\alpha'}(\frac{g_s^2}{v}\tilde Q_1+\tilde Q_5). 
\ee
The energy above extremality then reads
\be
E= M-M^{ext} \frac{vR\tilde r_0^2}{2g_s^2\alpha'^2}\left(e^{-2\alpha_1}+e^{-2\alpha_5}+1\right).
\ee
As explained in the main text, we take the decoupling limit, by which $\alpha_5\to \infty$ while $\alpha_1$ is kept fixed. In this limit, the dimensionless energy reads as follows
\be
{\cal E}\equiv R E = \frac{vR^2 r_0^2}{2g_s^2\alpha'^2}\left(e^{-2\alpha_1^+}+1\right)=\frac{\tilde Q_1}{\lambda \sinh2\alpha_1^+}\left(e^{-2\alpha_1^+}+1\right)\label{eq:dimE}
.\ee
Note we have rewritten the horizon radius in terms of $\alpha_1$ by means of \eqref{eq:BHparameters} and also introduced the deformation parameter $\lambda=\alpha'/R$.

Finally, the entropy associated to the solution is obtained by computing the area spanned by the $S^3$ at the horizon $r=r_0$. From \eqref{eq:5d-metric} one easily obtains
\be
S= \frac{{\rm Area}(S^3)}{4G_N^{5d}}= \frac{2\pi vR r_0^3 }{g_s^2\alpha'^2}\cosh\alpha_1\cosh{\alpha_5}\approx
\frac{2\pi vR \sqrt{\tilde Q_5} r_0^2}{g_s^2 \alpha'^{3/2}}\cosh\alpha_1
\ee
where, in going to the second equality, we have taken the decoupling limit. Finally, inverting the relation \eqref{eq:dimE} and plugging it into the above expression, one gets the equation of state \eqref{eq:entropy} for the entropy in terms of the energy above extremality.

\bibliographystyle{JHEP}
\bibliography{TTbarBib}

\end{document}